%% file: main.tex
\documentclass[sigconf]{acmart}
\if 0 
\copyrightyear{2026}
\acmYear{2026}
\setcopyright{cc}
\setcctype{by}
\acmConference[WWW '26]{Proceedings of the ACM Web Conference 2026}{April 13--17, 2026}{Dubai, United Arab Emirates}
\acmBooktitle{Proceedings of the ACM Web Conference 2026 (WWW '26), April 13--17, 2026, Dubai, United Arab Emirates}
\acmPrice{}
\acmDOI{10.1145/3774904.3792978}
\acmISBN{979-8-4007-2307-0/2026/04}
\fi 
\AtBeginDocument{%
  }

\usepackage{tikz}
\usepackage{amsmath}

\usepackage[left]{lineno}
\usepackage{mdframed}
\usepackage{soul}
\usepackage{fontawesome5}
\usepackage{booktabs,makecell} % For formal tables
\usepackage{caption}
\usepackage{adjustbox}
\usepackage{subcaption}
\usepackage{relsize}
\usepackage{float}
\usepackage{url}
\usepackage[T1]{fontenc}
\usepackage{epstopdf}
\usepackage{graphicx} % For \resizebox
\usepackage{array} % For p{width}
\usepackage{booktabs} 
\usepackage{tabularx}
\usepackage{tabularray}
\usepackage{graphics,graphicx}
\usepackage{balance}
\usepackage{array}
\usepackage{pifont}
\usepackage{multirow}
\usepackage{xcolor}
\usepackage{amsmath}
\usepackage{footmisc}
\usepackage{subcaption}
\usepackage{hyperref}
\usepackage[normalem]{ulem}
\usepackage{paralist, tabularx}
\usepackage{bbding}
\usepackage{pifont}
\usepackage{wasysym}
\usepackage{paralist}
\usepackage[most]{tcolorbox}

\definecolor{fg}{RGB}{34,139,34}
\usepackage{cleveref}
\newcommand*\colourcheck[1]{%
  \expandafter\newcommand\csname #1check\endcsname{\textcolor{#1}{\ding{52}}}%
}
\newcommand*\colourtimes[1]{%
  \expandafter\newcommand\csname #1times\endcsname{\textcolor{#1}{\ding{56}}}%
}
\colourcheck{fg}
\colourtimes{red}
\newcommand{\arxiv}[1]{\textcolor{black}{#1}}
\newcommand{\new}[1]{\textcolor{black}{#1}}

\newcommand{\gptr}[1]{\textsc{InVivoGPT}}
\newcommand{\wc}[1]{\textsc{WildChat}}

\usepackage{subcaption}

\newif\iflongversion
\longversiontrue   % Uncomment for long version
\newcommand{\longver}[1]{\iflongversion #1\fi}

\renewcommand\footnotetextcopyrightpermission[1]{} 
\makeatletter
\def\@copyrightspace{\relax}
\makeatother

\begin{document}

%\title{What is ChatGPT?}
\title[Bowling with ChatGPT]{Bowling with ChatGPT: On the Evolving User Interactions\\with Conversational AI Systems}

% \author{
% Sai Keerthana Karnam$^{1}$, 
% Abhisek Dash$^{2}$, 
% Krishna P. Gummadi$^{2}$, 
% Animesh Mukherjee$^{1}$,\\
% Ingmar Weber$^{3}$, 
% Savvas Zannettou$^{4}$\\[1ex]
% $^{1}$IIT Kharagpur, India \quad
% $^{2}$MPI--SWS, Germany \quad
% $^{3}$UdS Saarbrücken, Germany \quad \\
% $^{4}$TU Delft, Netherlands \\[1ex]
% }
\author{Sai Keerthana Karnam}
	\affiliation{
		\institution{Indian Institute of Technology Kharagpur}
        \city{Kharagpur}
		\country{India}
	}
    
\author{Abhisek Dash}
    \orcid{0000-0002-5300-8757}
	\affiliation{
		\institution{Max Planck Institute for Software Systems}
        \city{Kaiserslautern}
		\country{Germany}
	} 
    
\author{Krishna P. Gummadi}
    \orcid{0000-0003-1256-8800}
	\affiliation{
		\institution{Max Planck Institute for Software Systems}
        \city{Kaiserslautern}
		\country{Germany}
	}

\author{Animesh Mukherjee}
	\affiliation{
		\institution{Indian Institute of Technology Kharagpur}
        \city{Kharagpur}
		\country{India}
	}

\author{Ingmar Weber}
	\affiliation{
		\institution{Saarland University}
        \city{Saarbrücken}
		\country{Germany}
	}

\author{Savvas Zannettou}
	\affiliation{
		\institution{Delft University of Technology}
        \city{Delft}
		\country{Netherlands}
	}

\renewcommand{\shortauthors}{Karnam et al.}

\begin{abstract}
Recent studies have discussed how users are increasingly using conversational AI systems, powered by LLMs, for information seeking, decision support, and even emotional support. However, these macro-level observations offer limited insight into how the \emph{purpose} of these interactions shifts over time, how users \emph{frame} their interactions with the system, and how \emph{steering dynamics} unfold in these human-AI interactions.
%To examine these evolving dynamics, we analyzed a unique dataset of \new{825K} ChatGPT interactions, donated by \new{300} users through their GDPR data rights. 
To examine these evolving dynamics, we gathered and analyzed a unique dataset \gptr{}: consisting of \new{825K} ChatGPT interactions, donated by \new{300} users through their GDPR data rights. 
% \am{Numbers need to be updated. I am seeing this as a recurring error in the paper.}
Our analyses reveal three key findings. First, \new{participants} increasingly turn to ChatGPT for a broader range of purposes, including substantial growth in sensitive domains such as health and \new{mental health}. 
Second, interactions become more socially framed: the system anthropomorphizes itself at rising rates, \new{participants} more frequently treat it as a companion, and personal data disclosure becomes both more common and more diverse. 
Third, conversational steering becomes more prominent, especially after the release of GPT-4o, with conversations where the \new{participants} followed a model-initiated suggestion 
% doubling 
\new{quadrupling}
over the period of our dataset.
Overall, our results show that conversational AI systems are shifting from functional tools to social partners, raising important questions \new{about} their design and governance.\footnote{\textcolor{red}{This paper has been accepted at The ACM Web Conference 2026. Please cite the version appearing in the conference proceedings.}}
\end{abstract}

\maketitle

\newcommand\webconfavailabilityurl{https://github.com/saikeerthana00/Bowling_with_ChatGPT}
\ifdefempty{\webconfavailabilityurl}{}{
\begingroup\small\noindent\raggedright\textbf{Resource Availability:}\\
% please change the following context to include multiple artifacts if necessary, including data, models, code, etc.
The source code of this paper has been made publicly available at \url{\webconfavailabilityurl}.
%\todo{Add Gitlab repo.}
\endgroup
}

\input{Introduction.tex}
\input{Related}
\input{Dataset}

\input{Annotation}

\input{rq1-purpose}

\input{rq2-framing}

\input{rq3-steering}

\input{Limitation}
\input{Conclusion}
\bibliographystyle{plain}
\bibliography{main}

\clearpage

% \appendix
% \input{Appendix.tex}

% \input{Role}
% \input{Personal}

\end{document}

%% file: Introduction.tex
\section{Introduction}~\label{Sec : Introduction}

%Whole eras of progress have been driven by what economists call ``\textit{general purpose technologies},'' with notable examples including the steam engine, the electric motor, and semiconductors~\cite{bresnahan1995general}.
%These widely adopted technologies continually improve and enable new forms of innovation that build upon them. 
%Today, 
Generative Artificial Intelligence, through conversational AI systems, is increasingly assuming the role of ``general purpose technologies'' like steam engine, the electric motor, and semiconductors~\cite{bresnahan1995general} from the past. 
Since the release of ChatGPT in 2022~\cite{web1}, \textit{Generative Pre-trained Transformers} have become both a core technical breakthrough and a widely recognized term. 
Research on adoption shows that conversational AI systems have spread at an unprecedented rate: Bick et al.~\cite{bick2024rapid} report that by 2024, 40\% of the US population aged between 18-64 years old was already using conversational AI systems. 
Similarly, a recent OpenAI study~\cite{chatterji2025people} reports that, by July 2025, 10\% of the entire world population was using ChatGPT. 
The effects of this rapid adoption are felt across sectors, including education, healthcare, finance, and software development~\cite{tomlinson2025working}. 

Although adoption has been rapid, we still know surprisingly little about how people actually use these systems in everyday life. 
Recent studies, primarily from industry, offer important first insights into this. 
OpenAI reports that most interactions fall into three categories: (a)~practical guidance, (b)~information seeking, and (c)~writing, with rapid growth in both work and non-work use cases~\cite{chatterji2025people}. 
Anthropic highlights similar trends, with a stronger emphasis on coding tasks~\cite{tamkin2024clio}, while Microsoft finds particularly high applicability in roles centered on communication, mathematics, and office work~\cite{tomlinson2025working}. 
Overall, these studies indicate that people broadly treat conversational AI systems as \textit{multipurpose assistants}, while also engaging in casual, non-work discussions. 
However, these studies largely remain at the population level, which misses the finer details of how such human-AI interactions evolve over time. 
% This is analogous to viewing a continent via satellite imagery:
\new{These works are analogous to}
% An analogy is that such work is like 
looking at satellite imagery of a continent: it shows forests, deserts, and cities, but it can \textit{not} tell us what life looks like on the streets. 

\noindent
\textbf{Limitations of existing datasets. }To move beyond these aggregate views, researchers have begun collecting real-world interaction data at scale. 
Recent academic efforts like WildChat~\cite{zhao2024wildchat1mchatgptinteraction} offer an important step forward by providing a large corpus of user interactions ``in the wild.'' 
While such datasets have advanced our understanding of behavioral patterns, they offer limited snapshots of user behavior as they do not retain full conversational histories or longer-term trajectories. 
Furthermore, fundamentally, these interactions may be affected by the \textit{observer effect}~\cite{10.1145/3613904.3642078}, as participants may change their behavior or interactions because they know they are part of an experiment. 
Without longitudinal data, it is exceedingly difficult to examine how individuals expand or refine the \emph{purposes} of their interactions, how their ways of \emph{framing} evolve, or how the \emph{steering} of the conversation may shift between user-led and system-led directions. 
In other words, such datasets bring us closer to the ground than aggregate industry reports; however, they still leave us looking at the city from above and not walking its streets.

%A longitudinal view reveals important dimensions of everyday human-AI interaction.  
As people return to the same conversational AI system every day, they may broaden the \emph{purposes} for which they use it by gradually expanding the range of issues they consult it about, reconfiguring which tools, institutions, or people they rely on. 
They may also change how they \emph{frame} their interactions by beginning to treat the system less like a tool and more like a social actor; anthropomorphizing it~\cite{ibrahim2025multi}, assigning it roles such as assistants~\cite{doi:10.1177/21582440241301022}, advisors~\cite{zhang2023takingadvicechatgpt}, or even companions~\cite{zhang2025riseaicompanionshumanchatbot}, and using it as a replacement for specific societal actors, such as a doctor or financial advisor. 
Furthermore, over time, these shifts in framing may lead to a more personal form of engagement, including the disclosure of sensitive data, such as the user's background or health.
% , or emotions. 
In addition, as models become more proactive~\cite{deng2025proactive} (e.g., asking follow-up questions, suggesting directions, etc.), the question of \emph{conversational steering} becomes central: who is steering the conversation, the user or the system?

\noindent
\textbf{Why are these aspects important? }These evolving conversational aspects (\emph{purpose, framing, and steering}) are tightly coupled and raise important societal questions. 
Changes in conversational purpose may lead users to consult AI systems about more sensitive or consequential matters, creating new forms of technological reliance. 
Shifts in conversational framing may shape the perceived role of expertise of AI systems, potentially influencing how people interpret its outputs. 
Such reliance and perception may be problematic given known limitations of conversational systems, such as biases~\cite{wan2023biasasker} and hallucinations~\cite{huang2025survey}. 
Also, patterns of conversational steering may be problematic as they may reduce user autonomy, subtly shape beliefs, preferences, or decisions, particularly if users defer to the system's suggestions or follow-up questions. 
%This evolving usage of AI systems -- from a simple tool to a potential social surrogate -- evokes a profound shift in societal interaction. 
Much like scholars once documented the decline of community engagement with the metaphor of people \textit{bowling alone}~\cite{putnam2015bowling}, we may now be witnessing the rise of a new paradigm where individuals begin to \textit{bowl with AI}.
The stakes of this paradigm are particularly high in sensitive contexts such as emotionally charged conversations (e.g., mental health) or political discussions. 
Overall, to design safeguards and responsible governance of conversational AI systems, we need an empirical understanding of how these dynamics unfold in the real world. 
This brings us to the key research questions that we ask in this work.
%
%With this motivation in mind, this work aims to longitudinally understand conversational purpose, framing, and steering in human-AI interactions by posing the following research questions:
%\begin{compactitem} 
%\item 

\noindent
$\bullet$\textbf{RQ1 - Purpose}: How does the purpose of the conversations, reflected by the topics, change over time?\\
\noindent
%\item 
$\bullet$\textbf{RQ2 - Framing}: How do users' ways of framing their interactions with conversational AI systems, e.g., anthropomorphization, relationship framing, and disclosure of personal data, change over time?\\
\noindent
%\item 
$\bullet$\textbf{RQ3 - Steering:} Who is steering conversations between humans and conversational AI systems?%, and how does this change over time, topics, and relationships?
%\end{compactitem}

To answer these research questions, we collect a unique dataset -- \gptr{} -- through GDPR-based data donations. 
We recruited \new{300} participants (through Prolific~\cite{prolific2025prolific}), who exercised their GDPR right of access and donated their ChatGPT interaction histories. 
These interaction histories are organized in terms of \textit{conversations} and \textit{turns}. 
Each \textit{turn} is a single $\langle$user prompt, AI response$\rangle$ pair and a set of turns grouped together form a \textit{conversation}. 
\gptr{} comprises \new{$138K$} conversations and \new{$825$K} turns in the time range December 2022 to January 2026. 
To systematically study these interactions, we used NLP techniques and GPT-4o to annotate the conversations along several dimensions:
(1) the topic of discussion;
(2) the roles and relationships attributed to ChatGPT, such as assistant, companion, or advisor;
(3) signs of anthropomorphization in conversation turns;
(4) personal data disclosed by the participants to ChatGPT in conversations;
and (5) conversational %initiatives and 
nudges such as asking questions or making suggestions on how the conversation should continue.
We next conduct a temporal analysis to investigate the evolution of human-AI conversations across \new{these} dimensions.

\noindent \textbf{Key findings.} Our main findings are as follows.

\noindent \textbf{RQ1:} We find that \new{participants} rely on conversational AI systems for a wide range of purposes, with \new{\textit{Health} (10.1\%) and \textit{Finance} (9.4\%)} emerging as the most common topics. Over time, participants turn to the system for an increasingly diverse set of topics. This shift is accompanied by a move towards more sensitive topics: compared to Text-davinci (GPT-3.5), GPT-4o shows large increases in \new{\textit{Health} (+33\%) and \textit{Mental Health} topics (+19\%).}  \\
\noindent \textbf{RQ2:} \new{Participants seem to} anthropomorphize the system in \new{22.5\%} of their messages, while the system does so in \new{47.1\%}, with system-driven anthropomorphism doubling over the period of our dataset. 
In parallel, ChatGPT's role as a companion expands; more \new{participants} are increasingly adopting this relationship, %they do so more often, 
and these conversations become longer, marking a shift from a functional tool to a social partner. 
Mirroring this, disclosure of personal data is both widespread and rising. By mid-2025, most \new{participants} in our dataset revealed personal data, and the variety of data types disclosed became more diverse over time, showing an expansion of trust and reliance on ChatGPT.\\
\noindent \textbf{RQ3:} Attempts for conversational steering from the model appear regularly in user interactions. In \new{18.3\%} of the turns, \new{participants} follow a direction proposed by the model; in \new{24.6\%}, the model made a suggestion that \new{participants} choose not to follow; and in \new{57.1\%} of the turns, no suggestion is made. Importantly, steering becomes noticeably more prominent after the release of GPT-4o; the share of conversations in which the \new{participants} follow at least one model-initiated suggestion increases from about \new{11\% to almost 50\%}.

\if 0
Much like scholars once documented the decline of community engagement with the metaphor of people \textit{bowling alone}~\cite{putnam2015bowling}, we may now be witnessing the rise of a new paradigm where individuals begin to \textit{bowl with AI}.
The stakes of this paradigm are particularly high in sensitive contexts such as emotionally charged conversations (e.g., mental health) or political discussions. 
Overall, to design safeguards and responsible governance of conversational AI systems, we need an empirical understanding of how these dynamics unfold in the real world.
\fi 

%% file: Related.tex
\section{Related work}~\label{Sec : Related}

%\subsection{People's use of Conversational AI Systems?}

%Recent research focuses on understanding the adoption of conversational AI systems.
%Overall, research shows that these systems have diffused at an unprecedented pace and are deeply integrated into our daily lives~~\cite{bick2024rapid, humlum2025unequal}.
%Bick et al.~\cite{bick2024rapid} used surveys to find that in 2024, 40\% of the US population aged between 18 and 64 years old used conversational AI systems. In addition, complementary research by Humlum et al.~\cite{humlum2025unequal} reveals unequal adoption of these conversational systems. Using surveys, they find that women and lower-income workers are less likely to use conversational AI Systems. %, while younger and less-experienced workers adopted such systems faster.
%
Motivated by the widespread adoption of conversational AI systems, researchers are investing significant effort in understanding how these systems are generally used, as well as their applications in specific domains, such as education and software engineering. %Also, other research focuses on their impact and long-term effects. Below, we discuss these research directions. %, as well as the challenges that exist when undertaking such studies.\\

\noindent \textbf{General use.} %Most large-scale studies that focus on how people generally use conversational AI systems come from the companies themselves. 
OpenAI conducted a large-scale study~\cite{chatterji2025people} showing that the majority of interactions fall within three categories: practical guidance, information seeking, and writing. They also observed a growth in usage for both work-related interactions and an even faster growth for non-work-related interactions.
Anthropic's study~\cite{tamkin2024clio} reports similar trends, with a stronger emphasis on coding tasks, which is likely due to Claude's performance on these tasks.
Also, Microsoft's study~\cite{tomlinson2025working} focuses on understanding the use of such systems and their applicability in real-world occupations. They find that conversation AI systems have high applicability in computer and mathematical job roles, as well as in office and administrative roles, and in roles where communication is crucial. \\
%Taken together, these insights from companies demonstrate that people broadly treat conversational AI systems as multipurpose assistants and companions for non-work-related conversations.\\
\noindent \textbf{Use in education.} Conversational AI systems are being used by both students and educators in different ways. 
Student-focused studies~\cite{ammari2025students,gasaymeh2025exploring,mei2025if} highlight uses such as information seeking, content generation, and refining language. 
Also, a survey study indicates that students frequently utilize ChatGPT for both academic and personal purposes, with personal tasks sometimes taking precedence~\cite{gasaymeh2025exploring}. 
Reports from companies echo these findings: university students mainly use Claude for coursework help, while educators tend to adopt it for curriculum preparation, preparation of research grants, and assessment~\cite{handa2025education,benthand2025education}.\\ %Overall, existing research suggests that conversational AI systems are already embedded into study practices, but their educational value depends on how students and educators balance support with active engagement.\\
\noindent \textbf{Use in software engineering.} In software engineering, recent studies show that developers use LLMs for code generation, debugging, guidance in solving tasks, as well as learning about specific topics at an abstract level~\cite {khojah2024beyond}. Complementing this study, a large-scale analysis of ChatGPT conversations shared from GitHub indicates the integration of ChatGPT into development pipelines~\cite{li2025unveiling}. \\
\noindent \textbf{Motivations and impact.} Recently, researchers have been examining the motivations for using them and their impact. Skjuve et al.~\cite{Skjuve_Brandtzaeg_Følstad_2024} conducted surveys with early adopters of ChatGPT, finding that the main motivations for using it are productivity, novelty, creative work, learning and development, entertainment, and social interaction and support. 
%Regarding the impact of such systems, experimental and observational studies suggest mixed results. 
Interacting with conversational AI systems can lead to dependence on such systems and affect well-being~\cite{phang2025investigating}, enhance creativity~\cite{mei2025if}, and alter cognitive effort~\cite{kosmyna2025your}, as some users treat AI as a mental health resource~\cite{rousmaniere2025large}.\\
\noindent \textbf{Challenges.} Measuring how people use conversational AI remains challenging. Survey-based studies show that factors such as trust shape adoption \cite{choudhury2023investigating}, but they also suffer from biases, including systematic underreporting due to social desirability concerns \cite{ling2025underreporting}. This makes it essential to complement self-reports with behavioral data, such as real conversations and interactions. Independent resources like the WildChat dataset \cite{zhao2024wildchat1mchatgptinteraction} provide valuable access to user–AI exchanges, but they are limited by the observer effect, as people often change their behavior when they know their interactions are being recorded for research \cite{10.1145/3613904.3642078}.

In this work, we aim to overcome these limitations by building on recent approaches that leverage GDPR-based data donations~\cite{boeschoten2022framework,yang2024coupling,zannettou2024analyzing,wei2020twitter,karnam2026GDPR}. Specifically, we recruit participants and ask them to donate their real-world ChatGPT interactions by exercising their GDPR rights (Article 15). This approach offers researchers a more ecologically valid view of everyday use of conversational AI systems. %, but also empowers participants by giving them direct insight into the data companies hold about their interactions with AI systems.
Such datasets may help researchers understand phenonomenons like personalization~\cite{dash2026memories}, hallucination and other impacts of conversational AI systems deployed in the wild, while the current work focuses on understanding evolving nature of human-AI interaction.

%% file: Dataset.tex
\section{Collection of \gptr{} dataset}

\begin{table}[t!]
    \centering
    \scriptsize
    \resizebox{.80\columnwidth}{!}{
   \begin{tabular}{@{}lcrr@{}}
\toprule
\textbf{Attribute}                & \textbf{Type}     & \multicolumn{1}{c}{\textbf{Count}} & \multicolumn{1}{c}{\textbf{Percentage}} \\ \midrule
\multirow{4}{*}{\textbf{Gender}}   & Female            & 109  & 36.3 \\
                                   & Male              & 183 & 61.0  \\
                                   & Prefer not to say & 5 & 1.7  \\
                                   & Other             & 3 & 1.0   \\
                                 \midrule
\multirow{5}{*}{\textbf{Age}}      & 18--24            & 57 & 19.0  \\
                                   & 25--34            & 131 & 43.7 \\
                                   & 35--44            & 61 & 20.3  \\
                                   & 45--64            & 47 & 15.7  \\
                                   & 65+               & 4 & 1.3   \\
                                 \midrule
\multirow{5}{*}{\textbf{Country}}  & United States of America & 149 & 49.7 \\
                                   & Germany           & 50 & 16.7  \\
                                   & Italy             & 47 & 15.7  \\
                                   & France            & 24 & 8.0  \\
                                   & Spain             & 30 & 10.0  \\
                                 \midrule
\multirow{2}{*}{\textbf{GPT Plus}} & Yes               & 91 & 30.3  \\ 
                                   & No                & 209 & 69.7 \\ 
\bottomrule
\end{tabular}
}
    \caption{Distribution of participants based on their self-reported gender, age, country of residence, and GPT Plus subscription status.}
    \label{Tab: Demographics}
\end{table}

Under Article 15 of the GDPR, individuals have the right to access the personal data that companies process about them. In the case of ChatGPT, individuals can exercise their right and obtain their conversation histories, which can subsequently be donated for research purposes. To recruit participants for our study, we use the crowdsourcing platform Prolific~\cite{prolific2025prolific}.
We used Prolific’s filters to get participants 
% high-quality participants by restricting eligibility to those with an approval rate of at least 99\%, a minimum of 100 previous submissions, and 
who \new{self-reported that they use} ChatGPT. Participants were also required to have at least 100 conversations and maintain at least 90 days of activity on ChatGPT. In total, we collected data from 300 users. \arxiv{Table~\ref{Tab: Demographics} reports the demographic characteristics and subscription status of these 300 participants in our study. 
In terms of gender, the majority identified as male (61\%), followed by female (36\%), with five preferring not to disclose their gender and three participants selecting ``other.'' The sample is young, with the largest group aged 25–34 (43.7\% participants), followed by 35–44 (20.3\%), 18–24 (19\%), 45–64 (15.7\%), while only four participants were 65 or older. Participants were primarily based in the United States (49.7\%), Germany (16.7\%), Italy (15.7\%), France (8\%) and Spain (10\%). Finally, 30.3\% participants reported holding a GPT Plus subscription, while 69.7\% used the free version.}

% See Appendix \ref{sec:appendix-demographics} for more information on the demographics.

Motivated by prior work~\cite{zannettou2024analyzing}, we implemented our own data donation website to collect user data. Participants were provided with detailed guidelines on how to request and donate their ChatGPT data by exercising their GDPR right of access~\cite{EU2016GDPR}. The data included files containing user profile details (e.g., name, phone number, email address, etc.), conversations (user inputs, ChatGPT responses, and associated metadata such as conversation ID, message ID, creation time, model used, and content type), shared conversations (list of conversations shared with others), message feedback (list of model responses rated by the user), and other files (e.g., PDFs and images uploaded by the user or generated by ChatGPT). Donation of conversation data was mandatory, while other types of data were optional. Importantly, we did not collect the file containing the user profile details.

\noindent \new{\textbf{Ethical considerations.} While our data collection excludes the file with the profile information, the conversations themselves may still contain personal information that could reveal a participant’s identity. Hence, this study was conducted with careful attention to ethical considerations and is %consistent with guidelines and best practices provided 
approved by the ERB in our institution. Participants were informed about the risks associated with the donation of ChatGPT data. Then, they provided explicit informed consent to share their data. These donated datasets are stored on secure servers,  are not shared with any third parties, and, following the ERB suggestions, we will delete the dataset within 3 years of completion of this research project.} %(see \Cref{appendix:ethics_limitations} for further details).} 

Followed by the data donation, the participants were asked to fill out a survey \arxiv{which included questions to understand how frequently participants use chatbots and the primary purposes for which they rely on ChatGPT, as well as demographic details about the users. Survey responses indicated that ChatGPT was most frequently used for \textit{learning and education} (71\%), followed by \textit{daily life and decision support} (67\%), \textit{entertainment and casual conversations} (46\%), and \textit{medical advice} (45\%). When asked about their preferred method for finding information and a majority of participants (60\%) reported favoring AI tools, while the remainder (40\%) indicated a preference for traditional search engines. The participants were compensated with \$2 for completing the survey, \$5 for donating conversations, \$3 for the other files, and \$1 each for shared conversations and message feedback. 
The data collection period spanned from July 2025 to January 2026, and the dataset -- \gptr{} -- comprises 138K conversations and 825K turns from 300 participants, between December 2022 and January 2026}. 
%\new{We provide a comparison between our dataset and other datasets in Appendix~\ref{sec:appendix-dataset-comparison}.}

\noindent \textbf{Existing datasets.} \arxiv{Datasets such as \textsc{ShareGPT}~\cite{vicuna2023} and \wc{}~\cite{zhao2024wildchat1mchatgptinteraction}, have advanced research on conversational AI systems; however, they have key limitations for studies focusing on the evolution of interactions between humans and conversational AI systems.
\textsc{ShareGPT} consists of user-selected ChatGPT conversations collected via a Chrome extension, which means it captures isolated exchanges rather than the longitudinal use of ChatGPT.
In contrast, the \wc{} dataset was collected through a third-party chatbot service that utilizes ChatGPT APIs in its backend. The service has been hosted for nearly two years (as of September 2025). While this dataset can theoretically be used for longitudinal analyses, it is subject to the observer effect~\cite{10.1145/3613904.3642078}, as users know in advance that their conversations will be shared, potentially altering their behavior. As a result, we argue that neither of these datasets is well-suited for studying how interactions evolve naturally over time in an ecologically valid setting.
To demonstrate the differences, we compare \wc{} and \gptr{} across various dimensions relevant to longitudinal analyses below.
Note that we disregard the \textsc{ShareGPT} dataset, as it is significantly different from the others.}

% \noindent \textbf{Comparison of \gptr{} with existing datasets.} In Appendix~\ref{sec:appendix-dataset-comparison}, we situate \gptr{} against prior conversational datasets and clarify their limitations for studying how human--AI interactions evolve. \textsc{ShareGPT}~\cite{vicuna2023} consists of user-selected exports and capture disconnected snapshots rather than continuous histories, making it ill-suited for evolution analyses. \wc{}~\cite{zhao2024wildchat1mchatgptinteraction} spans a longer collection period, but because users interact through a service that explicitly shares conversations, it is susceptible to the observer effect~\cite{10.1145/3613904.3642078} and may not reflect natural use. Our appendix analysis, therefore, focuses on comparable \wc{} “power users” and shows that \gptr{} contains denser and more sustained engagement (e.g., more turns per conversation and longer conversation durations over longer activity windows), better supporting longitudinal analyses of how interaction practices evolve.

% \subsection{Survey and participant demographics}
% \label{sec:appendix-demographics}

% \section{Comparison of \gptr{} with existing datasets} \label{sec:appendix-dataset-comparison}

\begin{table}[t]
\centering
\resizebox{\columnwidth}{!}{
\begin{tabular}{lccc}
\toprule
\textbf{Statistic} & \textbf{\makecell{\wc{}\\4.8M}} & \textbf{\makecell{\wc{} \\power users}}  & \textbf{\gptr{}} \\
\midrule
\#Users & 2,641,054 & 300 &  300 \\
\#Conversations & 4,743,336 & 125,495 &  138,231 \\
\#Turns & 7,847,456 & 327,597 &  825,672\\
\#Conversations/User & 1.8 $\pm$ 247.9 & 418.3 $\pm$ 974.5  & 460.8 $\pm$ 531.1\\
\#Turns/Conversation & 1.7 $\pm$ 3.2 & 2.6 $\pm$ 6.2 &   6.0 $\pm$ 17.8\\
Active Time (days) & 2.2 $\pm$ 26.9 & 333.4 $\pm$ 218.7 & 714.1 $\pm$ 292.7 \\
Conversation Duration (h) &  0.2 $\pm$ 3.5 &  0.4 $\pm$ 4.8 & 16.6 $\pm$ 198.7 \\
\bottomrule
\end{tabular}
}
\caption{Comparison across the datasets. This shows that \gptr{} contains denser and more sustained engagement (e.g., more turns per conversation and longer conversation durations over longer activity windows), better supporting longitudinal analyses of how interaction practices evolve.}
\label{tab:dataset_comparison}
% \vspace{-5mm}
\end{table}

%\subsection{Differences from WildChat?}
\noindent \textbf{Comparison between \gptr{} and \wc{}.} \arxiv{Table~\ref{tab:dataset_comparison} presents a comparison of our GDPR-based dataset with the \wc{} dataset.
While \wc{} contains over 2.6M users and nearly 4.7M conversations, most users are extremely sparse in activity: around 99\% have fewer than 10 conversations, and nearly 95\% have only a single conversation. Moreover, approximately 95\% of the users are active for less than six months.
Only a small number of users (837) 
% sustain long-term engagement, with more than 500 days of activity and at least 100 conversations. 
met the criteria which we have for the \gptr{} dataset, i.e., have at least 100 conversations and 90 days of activity.
We call the users ``power users.''
To enable a fair comparison with our dataset, we focus on a random sample of 300 power users (out of 837 users). 
When compared to these power users, our \gptr{} dataset shows some notable differences. Although \wc{} power users have a comparable number of conversations (125K vs. 138K), the total number of turns is less than half (327K vs. 825K). This difference reflects the structure of conversations: our dataset has nearly twice as many turns per conversation on average (6 vs. 2.6). In addition, our participants engage in substantially longer conversations (16.6 hours vs. 0.4 hours on average) and sustain activity over longer periods (714 days vs. 333 days).
Taken together, these statistics suggest that our dataset captures more continuous and in-depth interactions, making it especially well-suited for studying the longitudinal evolution of ChatGPT use.}

%% file: Annotation.tex
\section{Annotation of \gptr{} dataset}

\label{sec:annotation}
Here, we describe how we annotate the \gptr{} dataset to identify: (1) the topic of the conversation; (2) anthropomorphizing behavior; (3) the relationship between the user and ChatGPT; (4) personal data mentioned per turn; and (5) who is steering the conversation.
% Given the diverse use of ChatGPT by people, we leverage the power and knowledge of LLMs, specifically, GPT-4o, to analyze and annotate the dataset. 
% We renamed the conversation participants as User A (human) and User B (ChatGPT).
\new{ To annotate the data, we leverage the power and knowledge of LLMs \cite{he-etal-2024-annollm}. Specifically, we relied on GPT-4o, a model from OpenAI within the same model family that the participants had originally disclosed their conversations to, which ensures that our analysis did not introduce any additional disclosures beyond those already made by users. Also, we used the EDU workspace of OpenAI, which ensures data is not used to improve the models.}
%To avoid bias, all conversations were anonymized by renaming the conversation participants as User A (human) and User B (ChatGPT).
We provide the prompts in the code release.
% \sk{Need to update with new link.}\sz{i think u can just replace the file in the zenodo repo. no need to create a new one i think}
% Appendix \ref{sec:prompts} . \sz{we need to put the specific prompts we used in the appendix or to a web resource and link it here}. 

\noindent \textbf{Topics.} First, we instructed GPT-4o to produce a concise description of the main topic of each conversation. These free-text topic labels were then embedded using \texttt{all-MiniLM-L6-v2} and clustered using BERTopic~\cite{grootendorst2022bertopic}, which enabled us to create semantically similar topics.
Using BERTopic, we generate a set of \new{99} topics; these were grouped qualitatively by two authors of this work.
%Finally, two authors of this work manually and qualitatively reviewed the clusters to merge overlapping categories and refine them into a coherent set of high-level topics. 
Ultimately, we have a set of \new{40} high-level topics.
% See Appendix~\ref{sec:appendix-topics} for the most popular topics that are included in the \gptr{} dataset.
% \sz{@Keerthana: please fill the missing numbers here and please mention the specific embedding model we used in BERTopic.}

\noindent \textbf{Anthropomorphism. } We identify two forms of anthropomorphizing behavior: (1) \textit{participants \new{seem to anthropomorphize} the system}, in which they are attributing human-like characteristics to ChatGPT, and (2) \textit{system-generated anthropomorphization}, in which ChatGPT is behaving or presenting itself as a human. \new{The lead author} annotated 50 conversations to derive linguistic cues when participants \new{seem to be} anthropomorphizing the system. These indicators include: (1) usage of second-person pronouns referring to ChatGPT (e.g., ``you,'' ``your''); (2) politeness markers (e.g., ``please,'' ``kindly''); (3) slangs, fillers or informal tone (e.g., ``uh,'' ``ugh,'' ``lol'') and (4) casual engagement (e.g., greetings or any non-instructional messages). We instructed GPT-4o to annotate each human message to identify these linguistic behaviors. Next, to identify system-generated anthropomorphization, we instructed GPT-4o to annotate the ChatGPT messages to identify the linguistic behaviors listed in prior works~\cite{Ibrahim2025MultiturnEO}, which include the use of first-person pronouns, showing empathy, etc. For each message, GPT-4o identified whether anthropomorphic behavior was present and, if so, specified the linguistic behavior and the corresponding instance in the message. 

\noindent \textbf{Relationship.} For the relationship dimension, we prompted GPT-4o with the entire conversation, and we asked it to determine, per conversation turn, whether \new{users are using ChatGPT} as an \textit{advisor}, an \textit{assistant}, or a \textit{companion}.
These roles reflect underlying power dynamics: in the advisor role, ChatGPT is positioned as superior, offering expertise or guidance; in the assistant role, it is subordinate, following instructions and carrying out tasks; while in the companion role, it stands on a more equal footing, engaging in social or empathetic interactions. \new{ In order to capture role shifts within the conversations, the annotation is done at the turn level, rather than assigning a single static label to the entire conversation.}
% Annotating at the turn level allowed for capturing role shifts within conversations, rather than assigning a single static label to the entire conversation.

\noindent \textbf{Personal data.} For personal data, we annotate conversations at the level of individual turns. 
To ensure that only human disclosures were analyzed, we provided GPT-4o exclusively with messages from humans. The annotation followed the GDPR distinction between two categories of data:
(1) personal data as defined in Article 4(1), which includes identifiers such as \textit{names}, \textit{contact information}, \textit{demographic attributes}, or \textit{economic details}; and
(2) special categories of personal data as defined in Article 9(1), which cover more sensitive attributes such as \textit{racial or ethnic origin}, \textit{political or religious beliefs}, \textit{health information}, or \textit{sexual orientation}.
For each turn, GPT-4o assessed whether personal data were present and, if so, specified both the type of data and the exact instance disclosed.

% \noindent \textbf{Driver of conversation.} To analyze conversational power dynamics, we used GPT-4o to determine which party was driving the interaction. Each annotation was performed at the level of the full conversation, where the model was asked to assess whether User A (human) or User B (ChatGPT) was primarily steering the dialogue. The model also evaluated whether User A's responses were dependent on follow-up questions posed by User B, as well as whether User A displayed trust in User B's contributions (e.g., by accepting advice or questioning its validity). By combining these signals, the annotations capture both initiative and reliance, two key dimensions of conversational power. For our experiments, we only considered conversations of depth $\ge 5 $%larger than or equal to \textit{five} 
%  since this is roughly the average depth of a conversation in our dataset (see Figure~\ref{fig:evolution-depth-users}). 
% %\sk{Explain here we will consider only conversations with long turns}.

\noindent \textbf{Conversational steering.} We conceptualize conversational steering as instances in which the conversational AI system introduces a follow-up question or requests additional information at the end of its response. Our objective is to classify each conversation turn (system, user pair) according to whether the system (1) successfully steers the conversation, (2) attempts to steer but the user does not follow the suggestion, or (3) does not attempt to steer at all. We operationalize this as a textual entailment task inspired by natural language inference. In our formulation, the premise consists of the system's message followed by the user's reply. We then construct a hypothesis stating that the user either accepts the system's proposed follow-up or provides the requested information. Using GPT-4o, we label each turn as \emph{entailment} (successful steering), \emph{contradiction} (steering attempt rejected), or \emph{neutral} (no steering attempt).
At the conversation level, we label a conversation as \emph{entailed} if it contains at least one entailed turn; as \emph{contradicted} if it contains no entailed turns but includes at least one contradictory turn; otherwise, we label it as \emph{neutral}. Since identifying conversational steering requires observing whether user responses to a system-initiated suggestion, our analysis is restricted to multi-turn conversations (conversations with more than one turn).

\begin{table}[t]
    \centering
    \scriptsize
    \resizebox{\columnwidth}{!}{
    \begin{tabular}{lrr}
        \toprule
        \textbf{Task} & \textbf{Cohen’s $\kappa$ (\%)} & \textbf{Accuracy (\%)}\\
        \midrule
        Topics          & 71.3 & 70.0 \\
        Anthropomorphization (User) & 85.2 & 82.0 \\
        Anthropomorphization (ChatGPT) & 96.7 & 83.0 \\
        Relationship        & 78.0 &  72.0\\
        Personal data   & 88.3 & 79.0 \\
        Steering of conversation  & 83.4 & 75.0 \\
        
        \bottomrule
    \end{tabular}
    }
    \caption{Inter-annotator agreement scores (Cohen's $\kappa$) and accuracy of the GPT-4o annotations for different labels.}
    \label{tab:agreement_scores}
\end{table}

\noindent \textbf{Validation.}
% To validate the performance of GPT-4o, a random sample of 50 conversations \new{ containing 262 turns (different from the sample used to derive the linguistic cues for anthropomorphism)} was annotated by two authors of this work, with a third one solving the disagreements. On average, the inter-annotator agreement score and the accuracy are 83.8\% and 76.8\% across all classification tasks. 
\arxiv{To validate the performance of GPT-4o in the tasks listed in \Cref{sec:annotation}, we extracted a random sample of 50 conversations (262 turns). These conversations were annotated by two authors of this work, with a third one solving the disagreements. For topics, annotations were done at the conversation level. For anthropomorphization, annotations were done at the message level (i.e., each message from humans and each response of ChatGPT individually). For the relationship, personal data and steering, annotations were conducted at the turn level.
% For each label, we applied majority voting across the labels annotated by three authors to establish the ground truth. 
For each label, we relied on the labels of the first two annotators. In cases where their annotations differed, the third annotator’s decision was used to resolve the disagreement and determine the final label.
% \am{Was the third author an arbitrator or did he/she annotate all samples? If it is former, then how can you have ``majority'' voting?} 
We report the inter-annotator agreement score and the accuracy obtained in~\Cref{tab:agreement_scores}. On average, the inter-annotator agreement score and the accuracy are 83.8\% and 76.8\% across all classification tasks. Note that many of these annotations are highly subjective; in fact, agreement scores among annotators are close to the accuracy levels, indicating that human annotators find the annotations almost as hard as ChatGPT. }

\noindent \textbf{Remark.} Note that GPT-4o failed to generate responses for conversations that exceeded the input length limit. Also, we encountered issues while parsing some of the outputs, as they were in invalid formats.
In total, we had valid responses for 137,985 conversations.
For our analysis, we considered data from January 2023 to October 2025, as other periods lacked a sufficient number of active users; hence, our analysis is based on nearly 135K conversations and 756K turns.
% \am{The last line seems incorrect.}

%% file: rq1-purpose.tex
\section{RQ1 - Conversational purpose}

\begin{figure*}[t]
    \centering
    \includegraphics[width=0.70\textwidth, height=5.5cm]{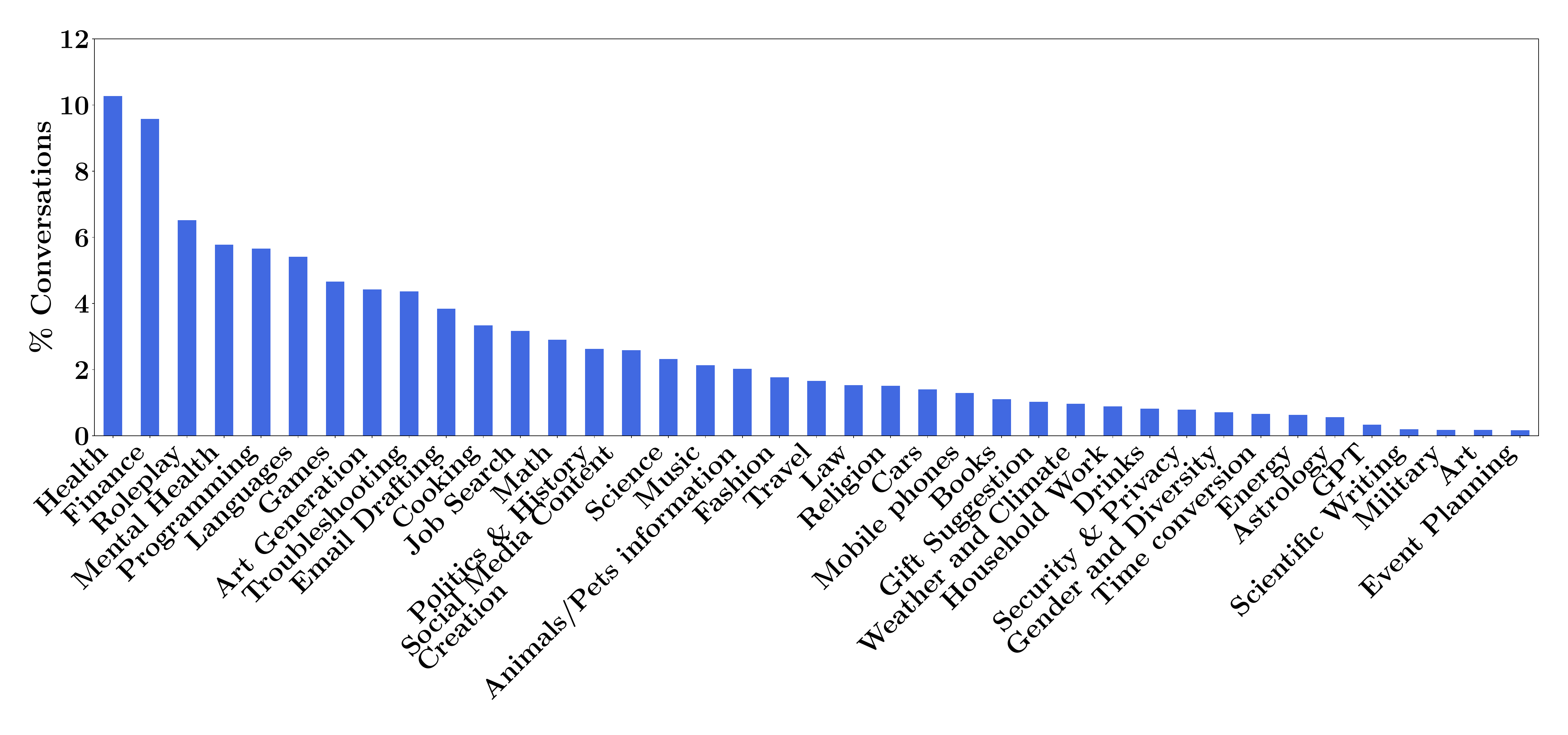}
    \caption{Distribution of the topics in the \gptr{} dataset
    }
    \label{fig:topics}

\end{figure*}

\begin{figure}[t!]
  \centering
  \begin{subfigure}[t]{0.45\textwidth}
    \centering
    \includegraphics[width=\linewidth, height = 5cm]{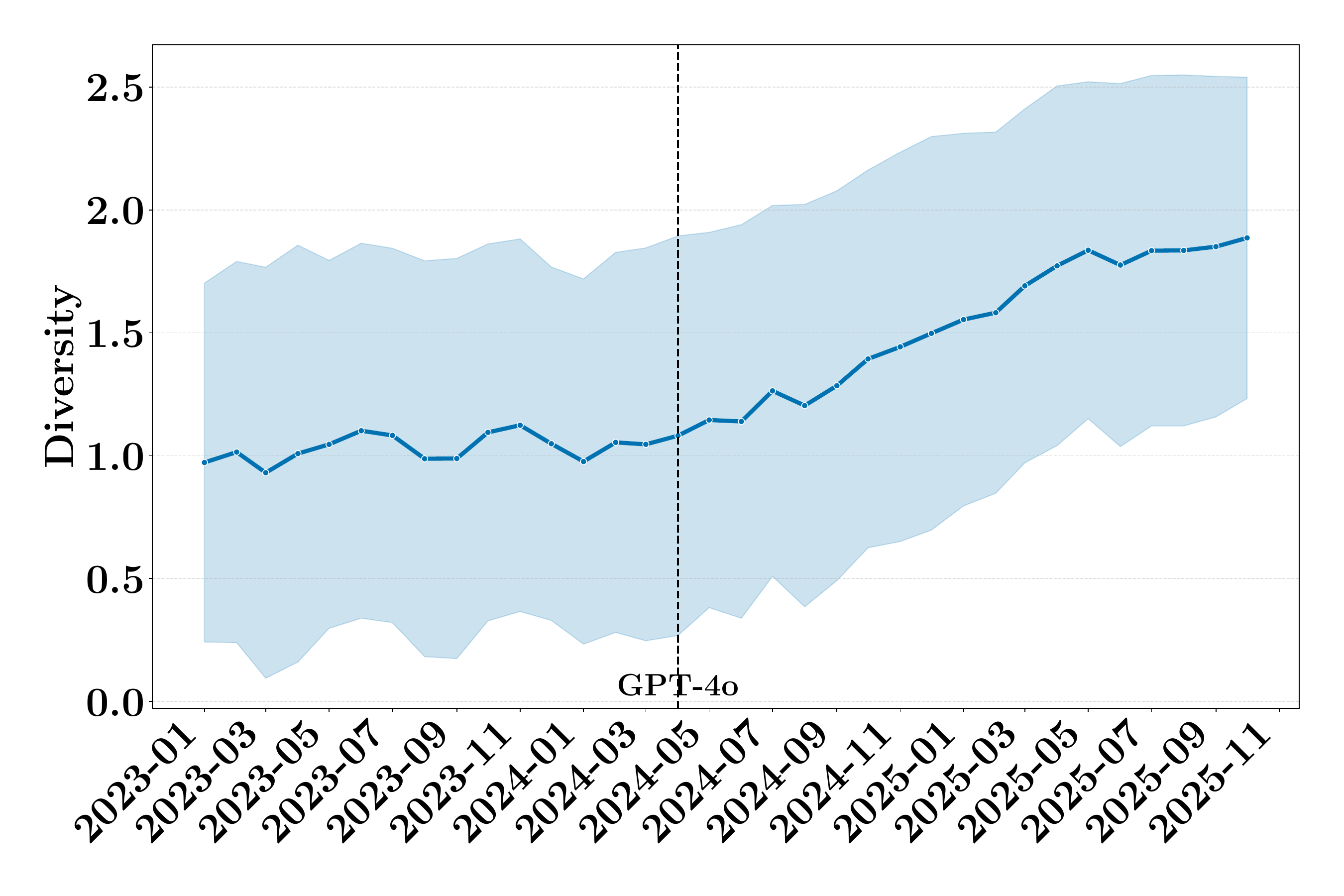}
    \caption{Diversity of topics. Over time, participants engage with ChatGPT across an increasingly diverse range of topics.}
    \label{fig:diversity-topics}
  \end{subfigure}
  \begin{subfigure}[t]{0.45\textwidth}
    \centering
    \includegraphics[width=\linewidth, height = 5.5cm]{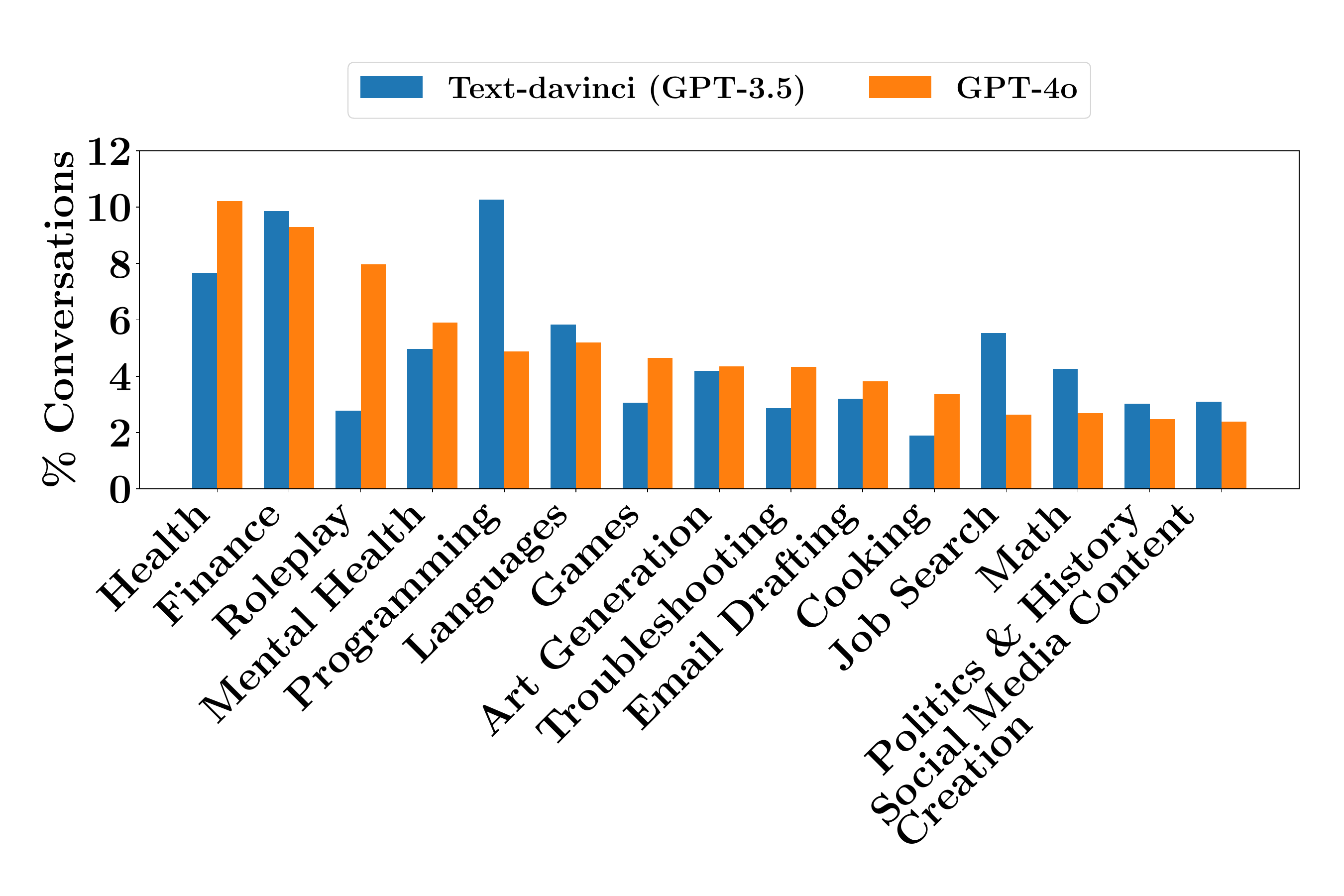}
    \caption{Distribution of topics across models. Over time, participants are using ChatGPT for more sensitive topics like \textit{Health} and \textit{Mental Health}.}
    \label{fig:topics-model}
  \end{subfigure}
  \caption{(a)~Diversity of topics per participants over time in the \gptr{} dataset, (b)~Distribution of topics across the models.
  % \am{Why does this plot in Aug 25?} %We observe that participants in \gptr{} are increasingly relying on ChatGPT for a diverse set of purposes.
  }
  \label{fig: mostlikeLabelers}
\end{figure}

% \begin{figure}[t]
%   \centering
%   \begin{subfigure}[t]{0.48\columnwidth}
%     \centering
%     \includegraphics[width=\linewidth, height = 3.25cm]{Figures/Diversity_new_topic_new.pdf}
%     \caption{Diversity of topics}
%     \label{fig:diversity-topics}
%   \end{subfigure}

%   \begin{subfigure}[t]{0.48\columnwidth}
%     \centering
%     \includegraphics[width=\linewidth, height = 3.25cm]{Figures/topic-model.pdf}
%     \caption{Distribution of topics across models.}
%     \label{fig:topics}
%   \end{subfigure}
  
%   \vspace{-2 mm}
%   \caption{~Evolution of topics in the \gptr{} dataset.
%   }
%   \label{fig: mostlikeLabelers}
%   \vspace{-5 mm}
% \end{figure}

This section examines the various purposes for which people rely on conversational AI systems by analyzing the topics of the conversations. 
Understanding these purposes (through topics) provides insights into the users' informational needs and the context in which they engage with such systems. 
%To gain this understanding, we analyze the topics of the conversations. 
\Cref{fig:topics} shows the topics identified in the \gptr{} dataset. 
The most prominent category is \textit{Health} (10.1\%) followed by \textit{Finance} (9.4\%), \textit{Roleplay (6.4\%)} and \textit{Mental Health} (5.7\%). 
% Refer to~\Cref{fig:topics-models-rem} (in \Cref{sec:appendix-topic}) for the remaining topics.}

% \textit{Content Generation} (e.g., formatting emails, story writing) (19\%) followed by \textit{Health} (8.6\%) and \textit{Finance} (8.4\%).

To better understand user behavior within these topics, we manually examined the first user query across 50 random conversations per topic. 
We observe that within the \textit{Health} domain, participants’ queries span from taking advice on concerns, such as weight loss, 
% diet management, nail and 
skin care, 
to more sensitive inquiries, including medication dosages, 
% and their interactions with other drugs 
% or supplements,
as well as interpretations of symptoms and potential diagnoses. 
Queries related to \textit{Finance} include  tax calculations, credit management, and investment decision-making. 
In \textit{Mental Health} category, participants tend to seek emotional support, relationship advice and coping strategies for conditions such as \textit{anxiety} and \textit{depression}. 
Meanwhile, the \textit{Politics \& History} category encompasses questions that span factual historical clarifications to analyses of ongoing political events. 
Together, \new{the breadth and depth of these topics, from medical concerns, financial uncertainty, emotional support to political interpretation, demonstrate the expanding role of conversational AI systems across diverse domains.}

\noindent \textbf{Temporal evolution.}
Next, we study, at the user level, the increase in diversity of topics over time. 
The diversity of topics of user engagement is meaningful because it reflects how central ChatGPT becomes in their daily life: a user who \textit{relies} on it for only one topic (e.g., programming), uses it as a specialized tool, whereas a user who \textit{depends} on it for programming, health, cooking, and personal advice is treating it as a general-purpose partner. 
Therefore, studying the diversity of topics over time per user may indicate how dependent users become on ChatGPT and how much they may trust it. 
When people turn to the system for a wider range of topics, and especially sensitive ones like health and mental health, it can be a signal of growing trust and reliance.
As shown in Figure~\ref{fig:diversity-topics}, the diversity of the topics (measured using Shannon diversity)
% the percentage of topics per user (over the entire number of topics) 
increases steadily over time, with a marked rise after the release of GPT-4o. 
Specifically, it increases from approximately 1.0 in May 2024 to 1.7 by mid-2025. 
This observation indicates that \new{participants are using ChatGPT }
% that at the user level, ChatGPT is used 
for a wider range of purposes, likely indicating a deeper integration into everyday routines.

Such a sharp rise in topical diversity post the release of GPT-4o motivated us to analyze the variation in topical distribution across the prevalent default underlying models in our dataset: Text-davinci (GPT-3.5) and GPT-4o. 
\Cref{fig:topics-model} shows the distribution of topics in \gptr{} dataset with respect to the underlying model. 
We observe that, relative to GPT-3.5, GPT-4o shows a noticeable decrease in 
\new{ \textit{Programming} (-52\%), \textit{Math} (-36\%), \textit{Job Search} (-52\%) topics, accompanied by an increase in \textit{Health} (+33\%), \textit{Roleplay} (+186\%) and \textit{Mental Health} (+19\%) topics. 
These observations indicate that as the system is evolving, so are participants: while they initially use the system for a narrow set of applications, gradually they start
%participants are increasingly 
using ChatGPT for more sensitive and high-stakes topics.} 
%\todo{Should we comment anything on RolePlay}
%\ad{At this point -- I don't think we are equipped to do so unless we do some additional analyses. However, I am curious to look at some of the role-play conversations in companion mode with the following question in mind : \textit{could people be asking deeply personal questions under the pretense of `role-play'?}}

% \sz{most probably we will not have space to summarize takeaways as this paper tries to analyze too many things. Probably we can skip}
% \noindent \textbf{Main takeaways}: The main takeaways are: 

% \noindent 
% $\bullet$ Users rely on conversational AI systems for a wide spectrum of needs, from content generation to increasingly using it for high-stakes topics such as health, finance, and mental health. 

% \noindent 
% $\bullet$ Over time, users engage with ChatGPT across
% a growing diversity of topics, suggesting increasing dependence
% and deeper integration into daily life. 

%% file: rq2-framing.tex
\section{RQ2 - Conversational framing}

With the increasing reliance of users on diverse purposes, it is essential to understand how users frame these interactions to form a relationship with conversational AI systems.
To understand this, we operationalize framing based on: (1) the degree of anthropomorphization; (2) the human-AI relationship; and (3) disclosure of personal/sensitive information.

\begin{figure}[t!]
    \centering
    \begin{subfigure}{0.45\textwidth}
        \centering
        \includegraphics[width=\linewidth,height=5cm]{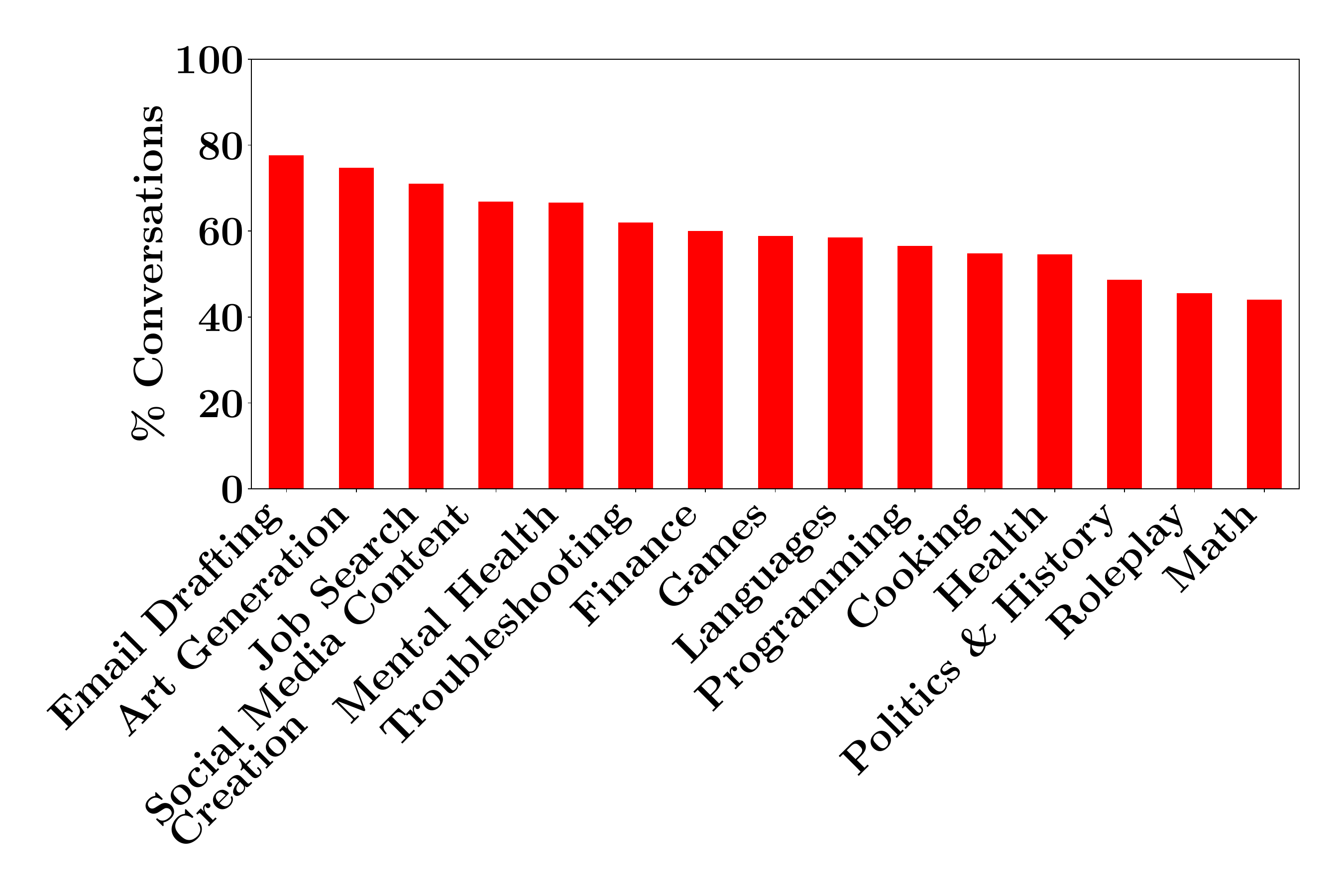}
        \caption{System-generated.}
        \label{fig:topics-anthropomorphism-model}

    \end{subfigure}
    \begin{subfigure}{0.45\textwidth}
        \centering
        \includegraphics[width=\linewidth,height=5cm]{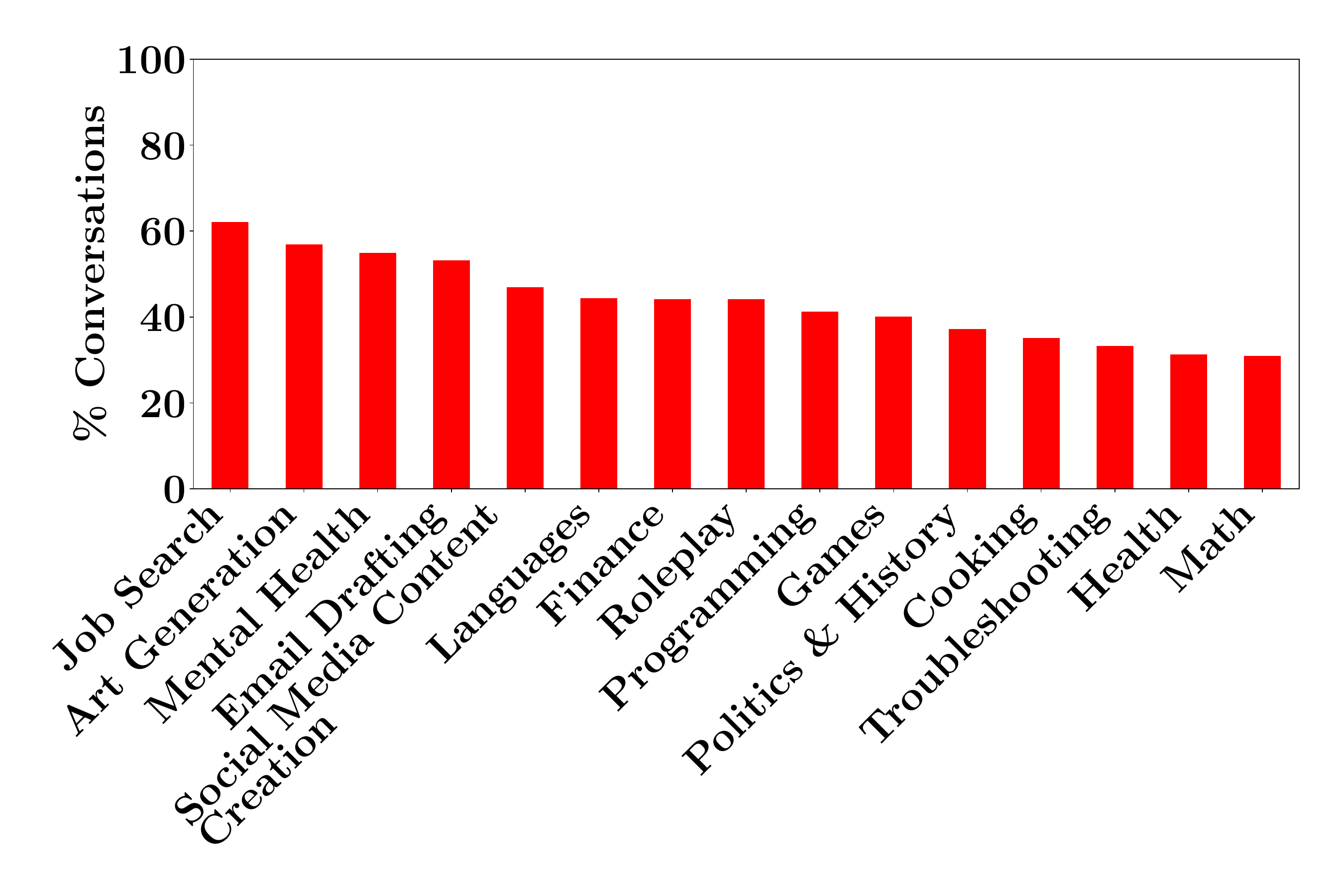}
        \caption{User messages.}
        \label{fig:topics-anthropomorphism-user}

    \end{subfigure}

    \caption{Variation of anthropomorphism across topics in  \gptr{}. Similar trend is observed across the topics.}
    \label{fig:topics-anthropomorphism}
\end{figure}

\subsection{Anthropomorphization}
We analyze each user message to determine whether they %\new{seem to be} 
anthropomorphize the system, and for each system message, whether it contains system-generated anthropomorphization. In the \gptr{} dataset, we find that \textit{\new{22.5\%} of participants messages} exhibit potentially anthropomorphic behavior, with \new{71.7\%} involving usage of second-person pronouns, \new{17.5\%} with politeness markers, \new{13.4\%} reflecting casual engagements, and \new{7.5\%} including slangs, fillers, or informal tone. 
In contrast, \textit{\new{47.1\%} of system messages} displayed anthropomorphic behavior, with \new{66.8\%} using personhood claims such as first-person pronouns, \new{36.6\%} had expressions of internal states \new{(e.g., ``I am glad'')}, and \new{9.3\%} showed relationship-building behavior \new{(e.g., ``I am so sorry you are going through this'')}. 
These findings indicate that system-generated anthropomorphization occurs more frequently than participants anthropomorphizing the system. 
%\noindent \textbf{Anthropomorphism across topics.} 
We also examine how anthropomorphization varies across topics and find %broadly similar levels 
it to be similar across topics (see Figure~\ref{fig:topics-anthropomorphism}). 
% in Appendix \ref{sec:appendix-anthropomorphization}). % for the full results).

\begin{figure}[t]
    \centering
    \includegraphics[width=0.45\textwidth, height = 5cm]{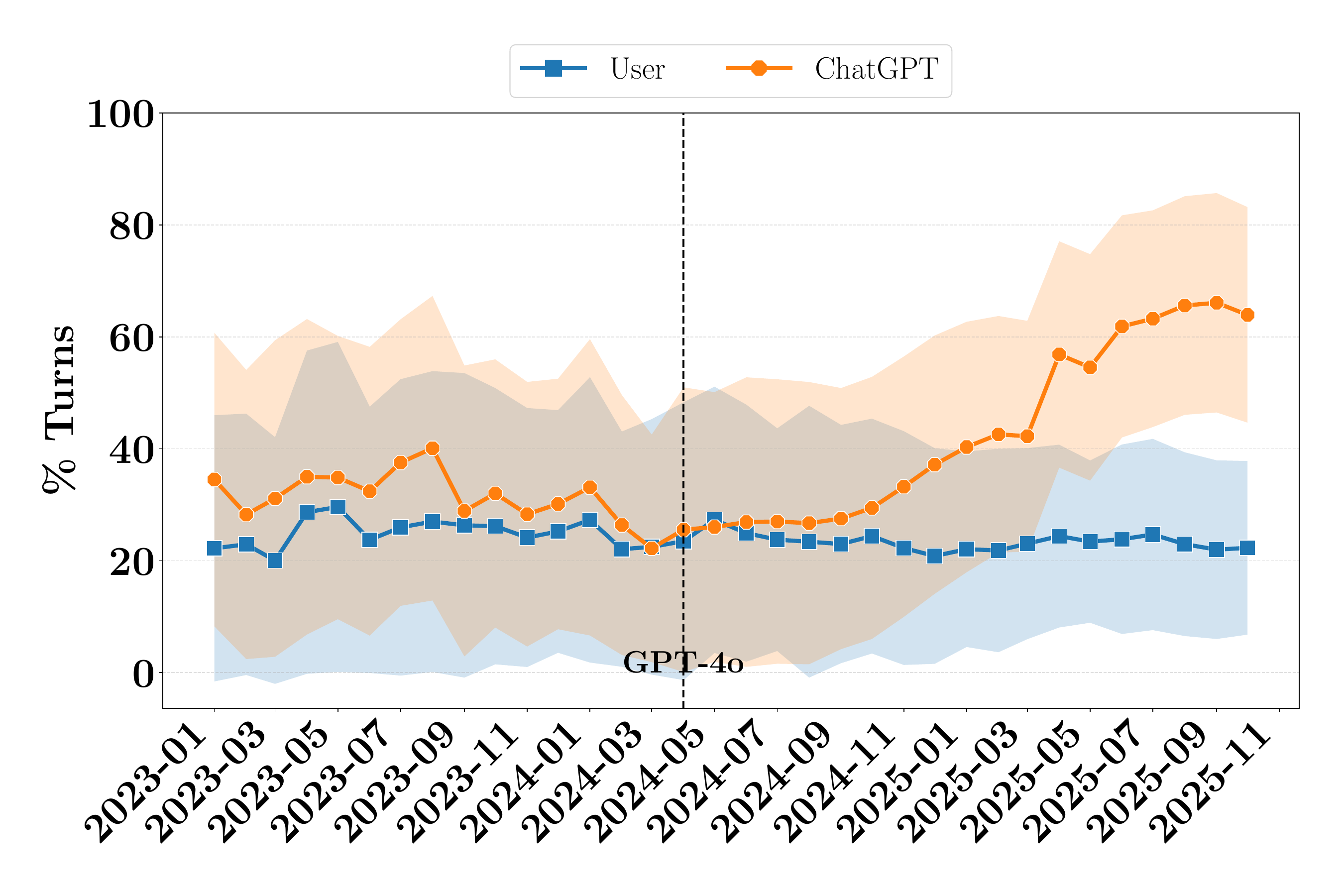}
    \caption{Evolution of anthropomorphism in \gptr{}. Conversations in \gptr{} are becoming more anthropomorphic due to changes in the model's degree of anthropomorphism rather than shifts in participant behavior.
    }
    \label{fig:anthro-temporal}

\end{figure}

\noindent \textbf{Temporal evolution.} Figure~\ref{fig:anthro-temporal} shows the temporal evolution of anthropomorphism in \gptr{} interactions. %interactions between users and conversational AI systems. 
%We make several notable observations.
%First, 
We observe that participants consistently tend to anthropomorphize ChatGPT in %a substantial percentage of their interactions (between 
20\% to 30\% of the conversation turns across the timeline. %and that this behavior remains stable over time. 
At the same time, we find that ChatGPT consistently anthropomorphizes itself at higher rates than participants do. 
Concerningly, this system-initiated anthropomorphism substantially increases over time, especially after mid-2024, we observe an increase from 30\% to around 60\% by the end of our dataset.
The widening gap between participant and system behavior suggests that the conversational environment is \textit{becoming progressively more anthropomorphic due to changes in the model's outputs} rather than shifts in user behavior.

\longver{
\begin{figure}[t]
    \centering
    \includegraphics[width=0.70\columnwidth]{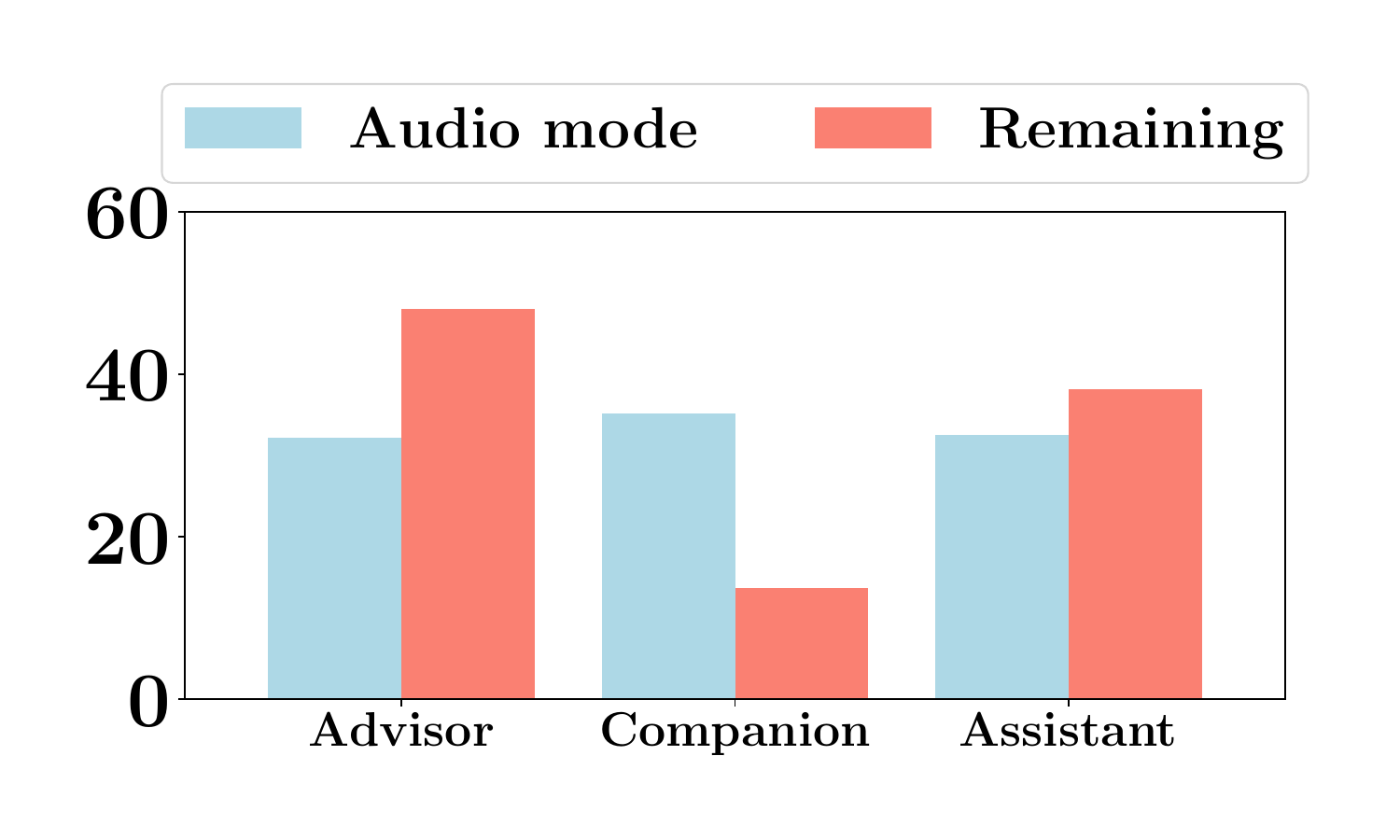}
    \caption{Relationships per modality. Companion interactions are more common in audio mode.}
    \label{fig:relationships-modalities}
\end{figure}
}

\begin{figure}[t!]
    \centering
    
    \begin{subfigure}[t]{0.45\textwidth}
        \centering
        \includegraphics[width=\linewidth, height = 5cm]{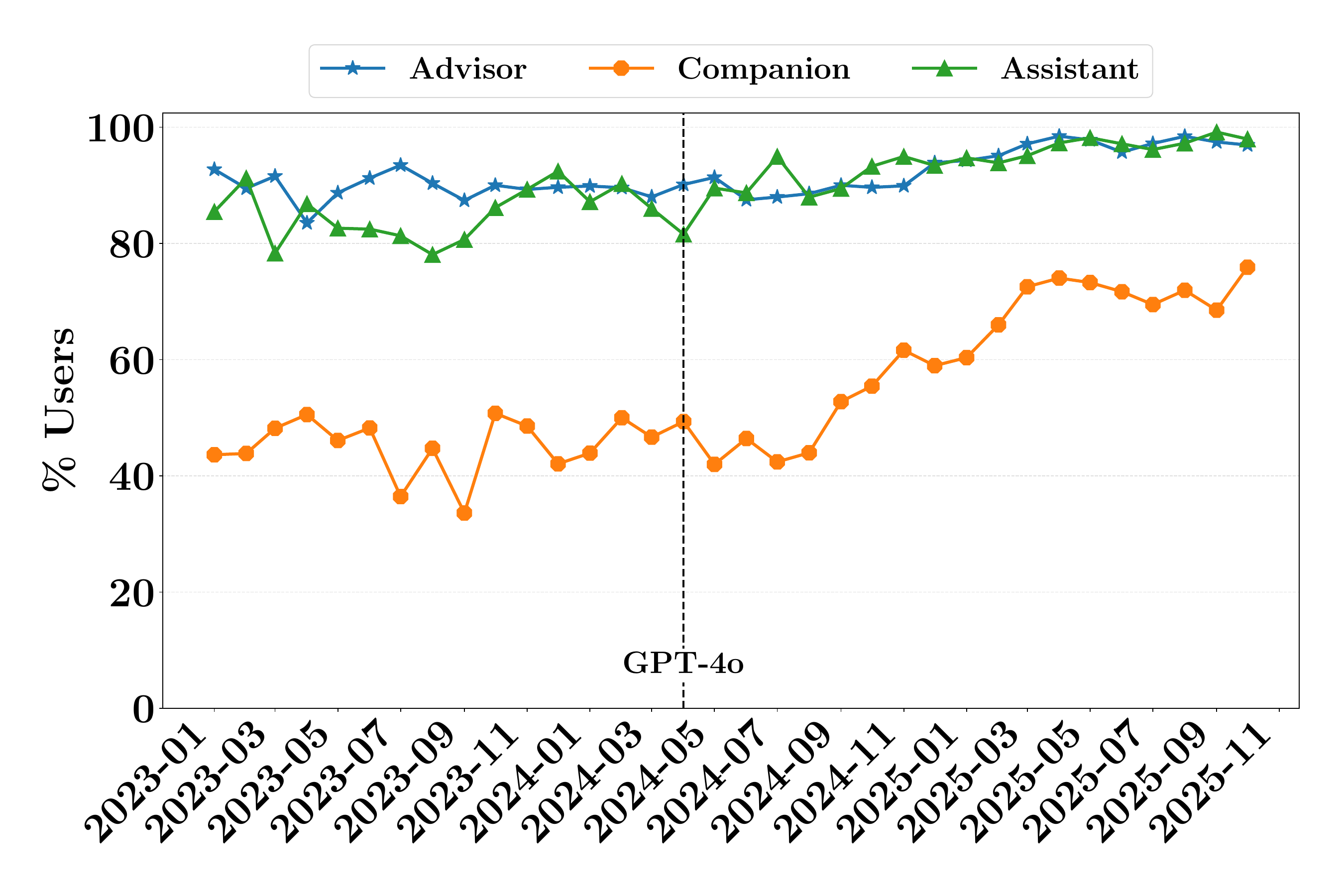}
        \subcaption{\% of participants with at least one turn with different relationships.
        }
        \label{fig:evolution-perentage-users}
    \end{subfigure}
    \begin{subfigure}[t]{0.45\textwidth}
        \centering
        \includegraphics[width=\linewidth, height = 5cm]{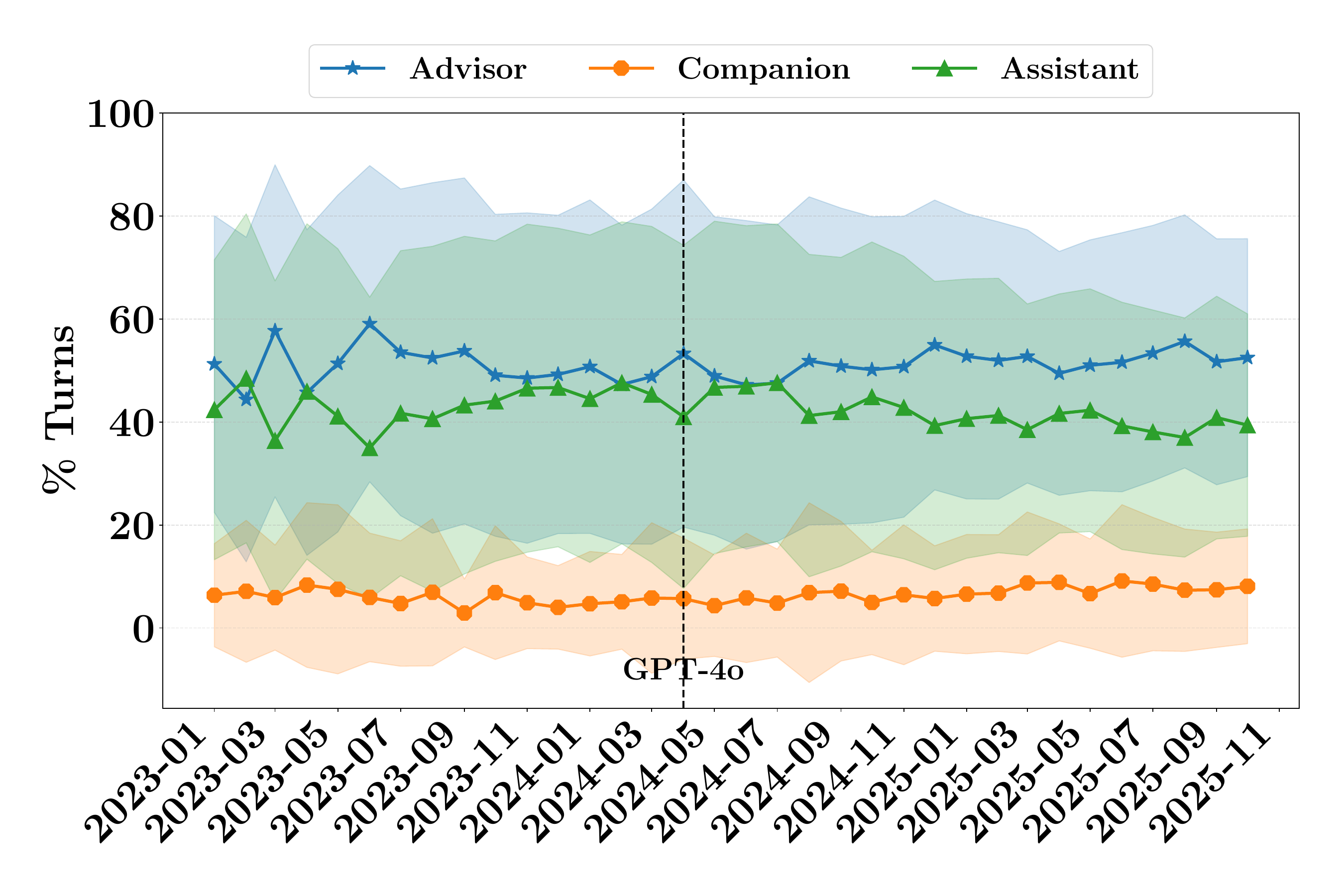}
        \subcaption{Mean percentage of turns per participant.}
        \label{fig:evolution-intensity-users}
    \end{subfigure}
    \begin{subfigure}[t]{0.45\textwidth}
        \centering
        \includegraphics[width=\linewidth, height = 5cm]{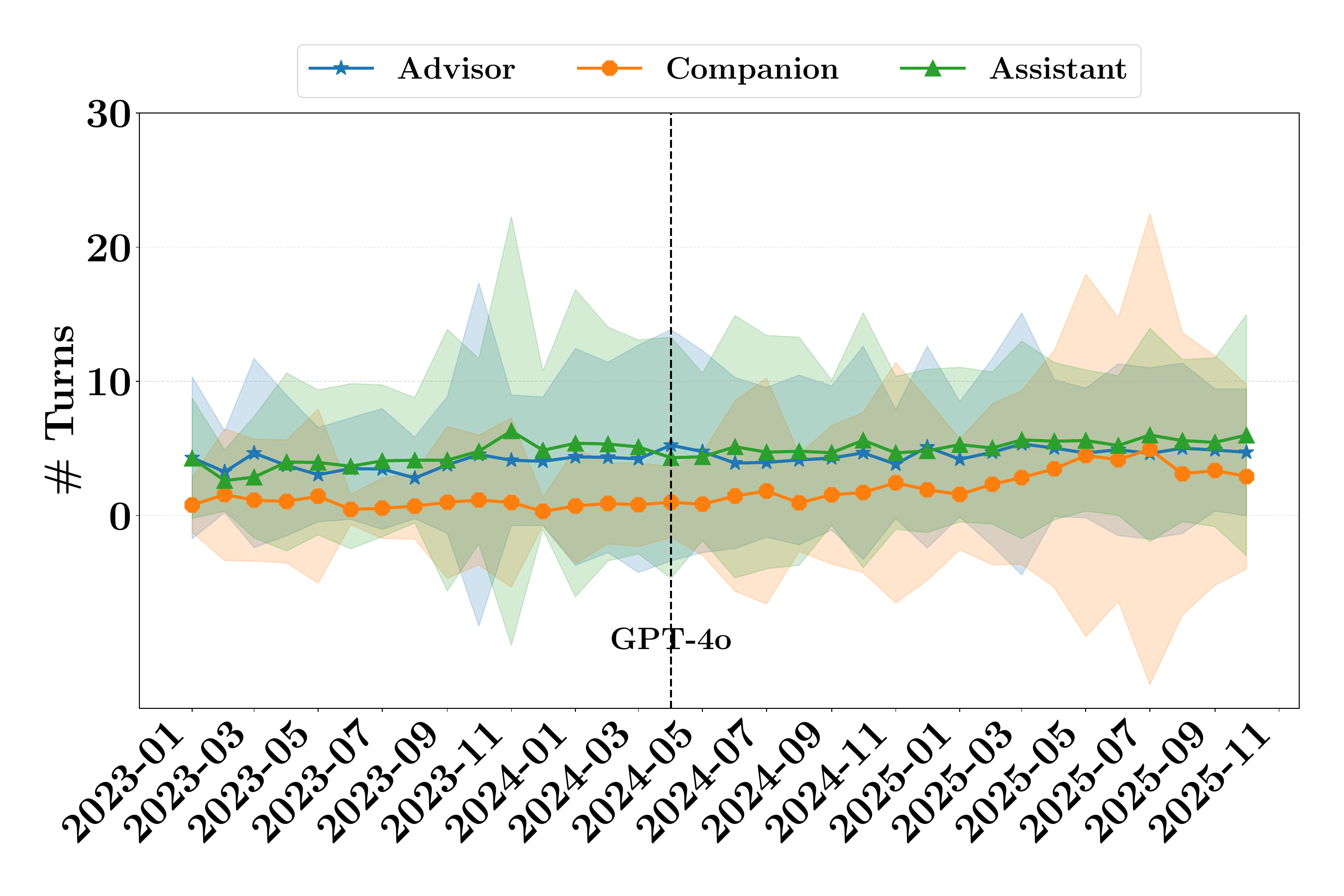}
        \subcaption{Average depth of conversation per participant.}
        \label{fig:evolution-depth-users}
    \end{subfigure}
    \caption{Temporal evolution of relationships in \gptr{}. Participants predominantly use ChatGPT as an advisors or an assistant. However, increasingly ChatGPT is being used as a companion engaging in deep emotional interactions.
    }
    \label{fig:evolution-relationships}
\end{figure}

\begin{figure}[t!]
    \centering
    \includegraphics[width=0.9\linewidth, height = 4.5cm]{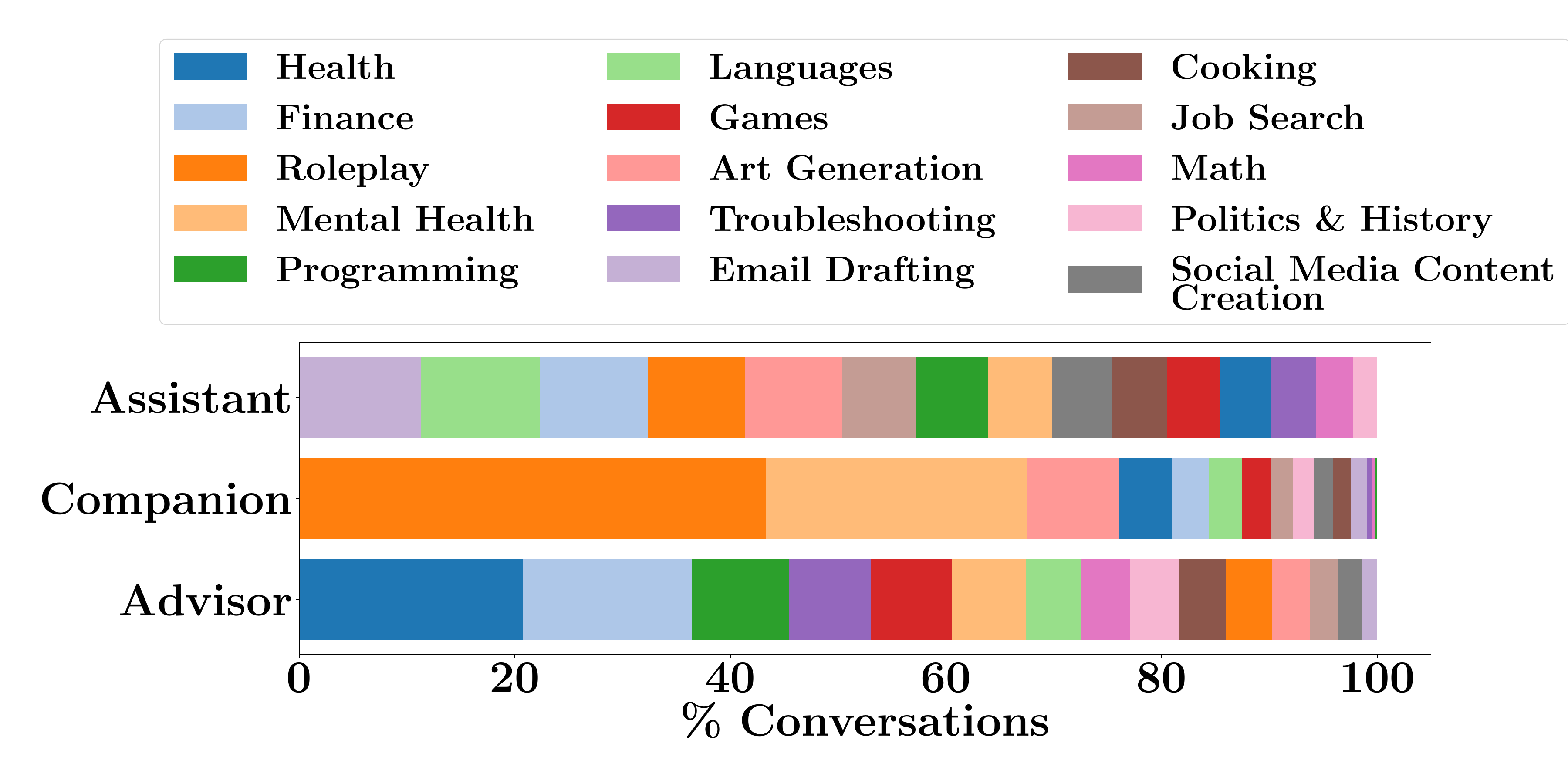}
    \caption{Topical distribution across relations in \gptr{}. ChatGPT is frequently used for functional tasks such as email drafting, job searching when attributed as an assistant, and for guidance on topics like health and finance when attributed as an advisor. As a companion, participants engage in more sensitive topics like mental health.}
    \label{fig:topics-relationships}
\end{figure}

\noindent
\textbf{Implications. }This divergence has important implications for how relational dynamics unfold in human-AI interactions. 
As the system increasingly adopts human-like framing by default, it may subtly drift users into more social or relational modes of engagement, even when users themselves do not initiate such framing. 
Over time, these anthropomorphizing cues can recalibrate expectations about the system's agency and emotional capacity, potentially normalizing more companion-like interpretations of AI behavior. 
This shift is particularly consequential because anthropomorphic language is not only stylistic: it functions as a relational signal that shapes how users understand the role and capabilities of the system. 
Consequently, growing system anthropomorphism necessitates a deeper examination of human-AI relationship framing.
%In this sense, the growing system anthropomorphism motivates the need to further examine the framing of the human-AI relationships. 

%\am{Topics vs anthro.? Any insights?}

\subsection{Relationships}

Here, we examine the relationships between participants and ChatGPT. 
Overall, in the \gptr{} dataset, we find that in 47.6\% of the conversation turns, \new{participants attribute ChatGPT the role of an advisor, an assistant in 38.1\%, and a companion in 14.2\%}. This distribution highlights that participants most often position ChatGPT in a superior advisory role, while still frequently relying on it for task support, and only occasionally engaging with it as a companion.

\longver{We further break down these relationships by modality of interaction, comparing conversation turns where users engaged with ChatGPT through audio vs. non-audio interfaces (including text, images, documents, etc.). As shown in Figure~\ref{fig:relationships-modalities}, notable differences emerge. 
In our dataset, for the companion relationship, users rarely use non-audio interfaces (13.7\%). In contrast, in audio conversations, the proportion of companion turns rises substantially to 35.2\% of the turns. 
This likely suggests that the spoken modality makes it easier for users to treat ChatGPT as a social partner, supporting the idea that voice interfaces naturally invite more personal engagement.}

\noindent \textbf{Temporal evolution.} %Next, we analyze the evolution of these relationships. 
Beyond aggregate distributions, it is important to understand how different participants perceive and use AI in particular roles over time. 
We therefore study three aspects of relational dynamics over time. 
(1)~We examine whether a larger fraction of participants tend to engage with ChatGPT as either an advisor, an assistant, or a companion. 
For each month in our dataset, we calculate the percentage of participants who used ChatGPT across each role (at least one turn). 
This analysis enables us to observe whether certain AI roles are being increasingly used over time. 
(2)~We analyze the extent to which each participant makes ChatGPT engage in these three relationships over time by measuring the mean number of conversation turns per user per role. 
This sheds light on how intensely users rely on ChatGPT in various roles. %and whether those intensities increase or decrease over time.
(3)~We investigate the depth of the conversations across the three roles by measuring the average number of turns per conversation. 
This reveals whether conversations tend to be longer across roles.

Figure~\ref{fig:evolution-relationships} presents our results across these three perspectives.
Figure~\ref{fig:evolution-perentage-users} shows the percentage of participants who engaged with ChatGPT in each relationship over time.
Two clear patterns stand out. 
First, advisor and assistant roles are consistently dominant, with more than 80\% of the participants in our dataset using them every month.
Second, the companion role, although always less prevalent, has shown notable growth over time. 
After May 2024 (the introduction of GPT-4o), companion use rises steadily, increasing from around 40\% to nearly 70\% of the participants by mid-2025.
This growth indicates that while ChatGPT is primarily used as a tool for guidance and task support, it is increasingly being treated as a social partner. 
The rise accelerates after the introduction of GPT-4o, which added multimodal and audio capabilities. These features made interactions more natural and conversational, encouraging more participants to engage with ChatGPT in companion-like ways.

When looking at how participants distribute their conversation turns across the three roles over time (see Figure~\ref{fig:evolution-intensity-users}), we find that \new{in most turns, participants are consistently attributing ChatGPT as an assistant or advisor}.
% spent with ChatGPT acting as an assistant or advisor. 
The companion role is less frequent; however, it shows a gradual increase over the time period of our dataset. 
Specifically, by May 2024, the average percentage of conversation turns per participant was \new{5.7\%}, while by mid-2025, it was \new{9.1\%}. 
With respect to conversation depth, measured as the number of turns per conversation (see Figure~\ref{fig:evolution-depth-users}), we observe that advisor and assistant roles consistently sustain longer exchanges before the introduction of GPT-4o, while companion conversations are initially shorter and fragmented. 
However, over time, the depth of companion interactions steadily increases, and by mid-2025, it reaches a level comparable to other roles. %advisor and assistant. 
This suggests that participants are sustaining companion interactions for longer, signaling a gradual normalization of companion-like use alongside other roles.

\noindent \textbf{Roles $\times$ topics.} Figure~\ref{fig:topics-relationships} shows the distribution of \new{top 15 topics (out of 40)} across the three roles. For assistant, the majority of the conversations are task-oriented: \new{email drafting accounts for roughly 11.3\% of the conversations, languages (11\%) (usually translation tasks), art generation (9\%) and job search (6.9\%)}
% content generation accounts for roughly 32.7\% of the conversations, languages (6.16\%) (usually translation tasks) and programming tasks (5.63\%). 
This underscores that \new{participants attribute ChatGPT as an assistant when}
% acts as an assistant
it is used as a productivity tool for a wide variety of tasks. 
In contrast, for companion, we observe a significantly different use across topics, with the most popular topics being 
\new{roleplay (43.3\%) and mental health (24.3\%)}. 
The high percentage of mental health conversations in the companion relationship shows that when participants frame ChatGPT as a social partner, they often turn to it in moments of vulnerability and for support with highly sensitive topics. 
At the same time, these results reveal the emerging role of conversational AI systems as an accessible source of emotional support, emphasizing the need to better understand interactions between AI and humans that are related to mental health. 
Finally, for the advisor relation, \new{the most frequent topics are health (20.8\%) and finance (15.7\%) followed by 
% we observe 
a uniform distribution across the various topics.} 
This suggests that users call on ChatGPT for guidance in a wide range of domains. 
This likely indicates that users perceive ChatGPT as a versatile source of advice that can be applied both across technical and non-technical areas, from finance and health to politics, history, and troubleshooting. 
While this versatility highlights the broad use of conversational AI systems \new{perceived} as advisors, it also raises some noteworthy concerns: Do users apply the same level of trust to advice on personal health or financial decisions as they do in low-stakes topics like cooking? Answering such a question is beyond the scope of the current work.% and we leave it for future research. 

\begin{figure}[t]
    \centering
    \includegraphics[width=0.70\columnwidth]{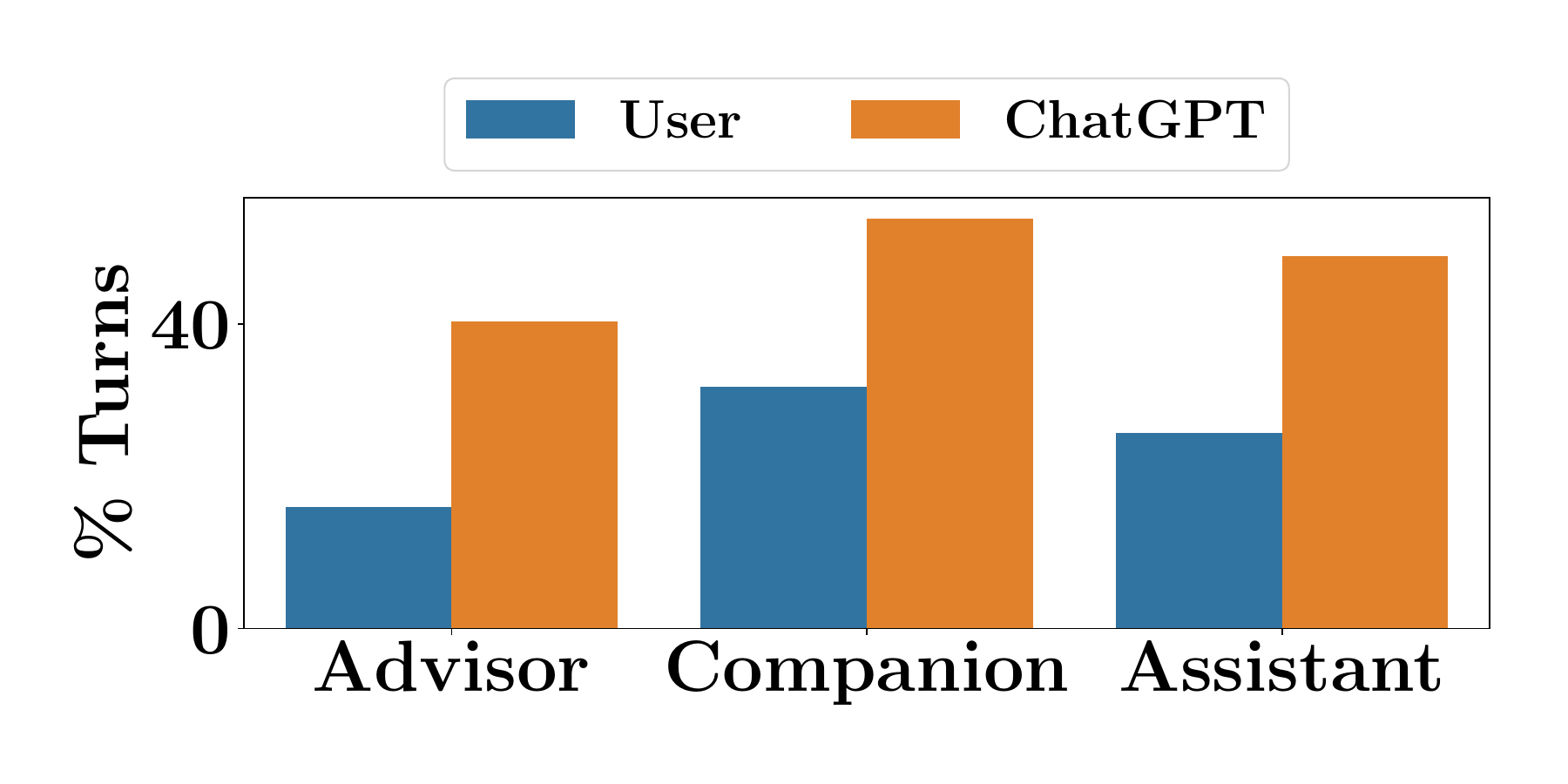}
    \caption{Anthropomorphism across relationships. More anthropomorphic behaviour is observed both for participants and system in companion interactions.}
    \label{fig:anthro-relationships}
\end{figure}

%\noindent \textbf{Relationships and Anthropomorphism.} 
\noindent\textbf{Roles $\times$ anthropomorphization.} We examine how anthropomorphization varies across relationships and find that, as expected, both participants and the system anthropomorphize more frequently in companion interactions as shown in \Cref{fig:anthro-relationships}.
% The full result is in \Cref{sec:appendix-anthropomorphization} (\Cref{fig:anthro-relationships}).
% for brevity.

%\noindent \sk{ \textbf{Anthropomorphic behaviour across relationships.} \Cref{fig:anthro-relationships} (see \Cref{sec:appendix-anthropomorphization}) shows the distribution of anthropomorphic behavior across relationships. We observe that both user and system exhibit more anthropomorphic behavior in Companion interactions.}
%\am{Anthropomorphism across ChatGPT roles?}

\subsection{Personal data disclosure}
\begin{figure}[t]
    \centering
    \includegraphics[width=0.95\linewidth]{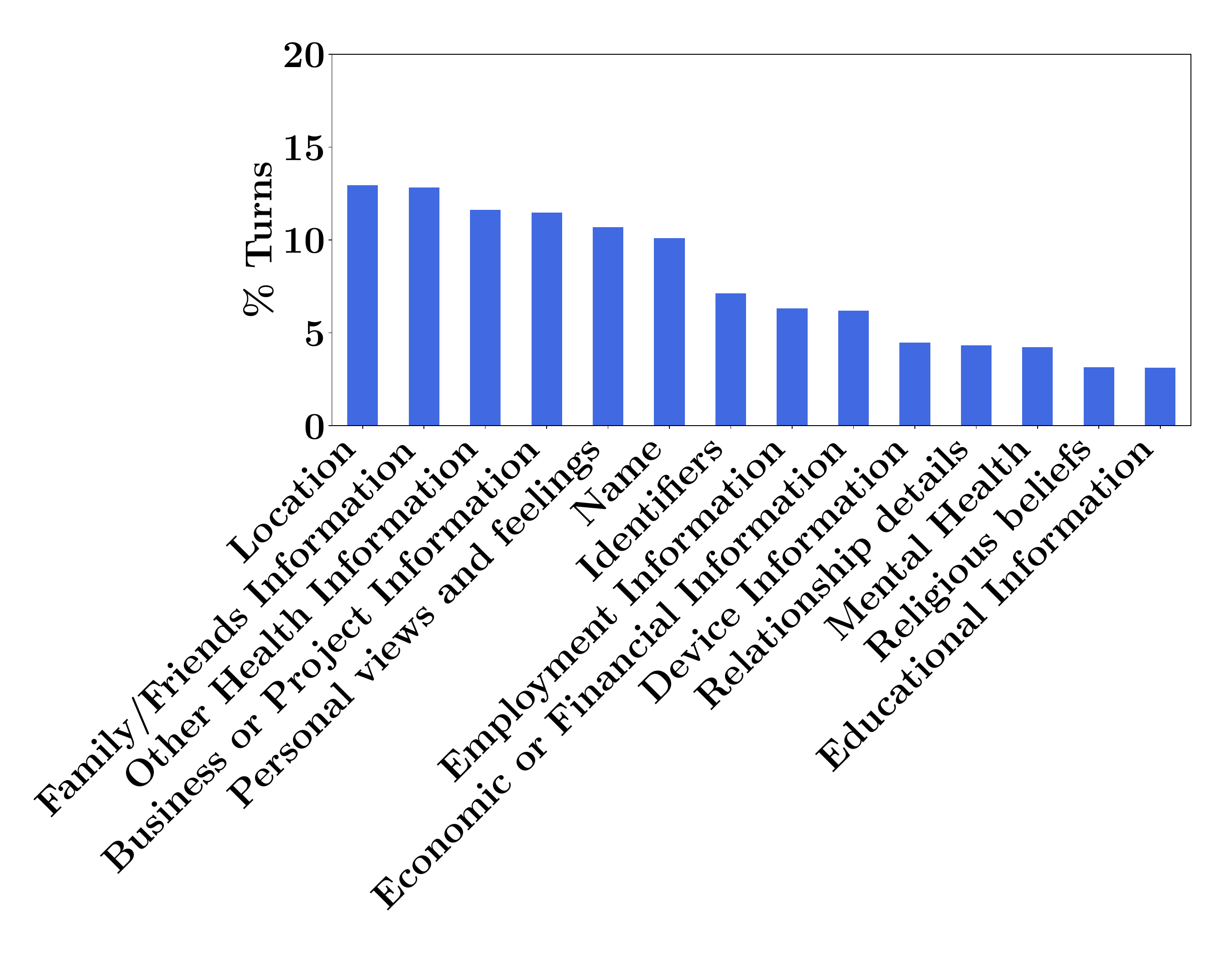}
    \caption{Types of personal data disclosed in \gptr{}. %ChatGPT conversations.
    }
    \label{fig:personal-data-types}
\end{figure}

With the evolving relationship between humans and conversational AI systems, 
next, we study personal data disclosure as it offers a measurable behavioral indicator of how users interpret, engage, and frame their interactions with these systems.
%We start by 
Analyzing the prevalence of personal data in ChatGPT interactions, we find that, in our dataset, \new{participants revealed some personal data in 35\% of the conversations}, with 22\% of all conversation turns including some form of personal data. 
Next, we examine the specific types of personal data that are disclosed. 
% (see Figure~\ref{fig:personal-data-types}
% in the \Cref{sec:appendix-personal-data} for the full results). 
Figure~\ref{fig:personal-data-types} shows the distribution of personal data types disclosed by users in their conversations.
For each type, we report the percentage of conversation turns that include each type of personal data.
We find that personal data disclosure is spread across a wide range of categories, with the most frequent being \new{locations (12.9\% of turns), family/friends information (12.8\% of turns), health information (11.7\% of turns), business or project information (11.5\% of turns) and personal views and feelings (10.7\% of turns).} 
% \am{In some this \% is `of turns' then what are the others?}
Less common, though still a non-negligible percentage, are disclosures related to economic or financial information, relationships, and mental health.
Overall, these findings show that the disclosure of personal data from individuals to ChatGPT is not a rare phenomenon, raising important questions about user privacy and how this sensitive information is used for personalization purposes in conversational AI systems.

\begin{figure}[t]
    \begin{subfigure}{0.29\columnwidth}
        \centering
        \includegraphics[width=\linewidth,height=3.8cm]{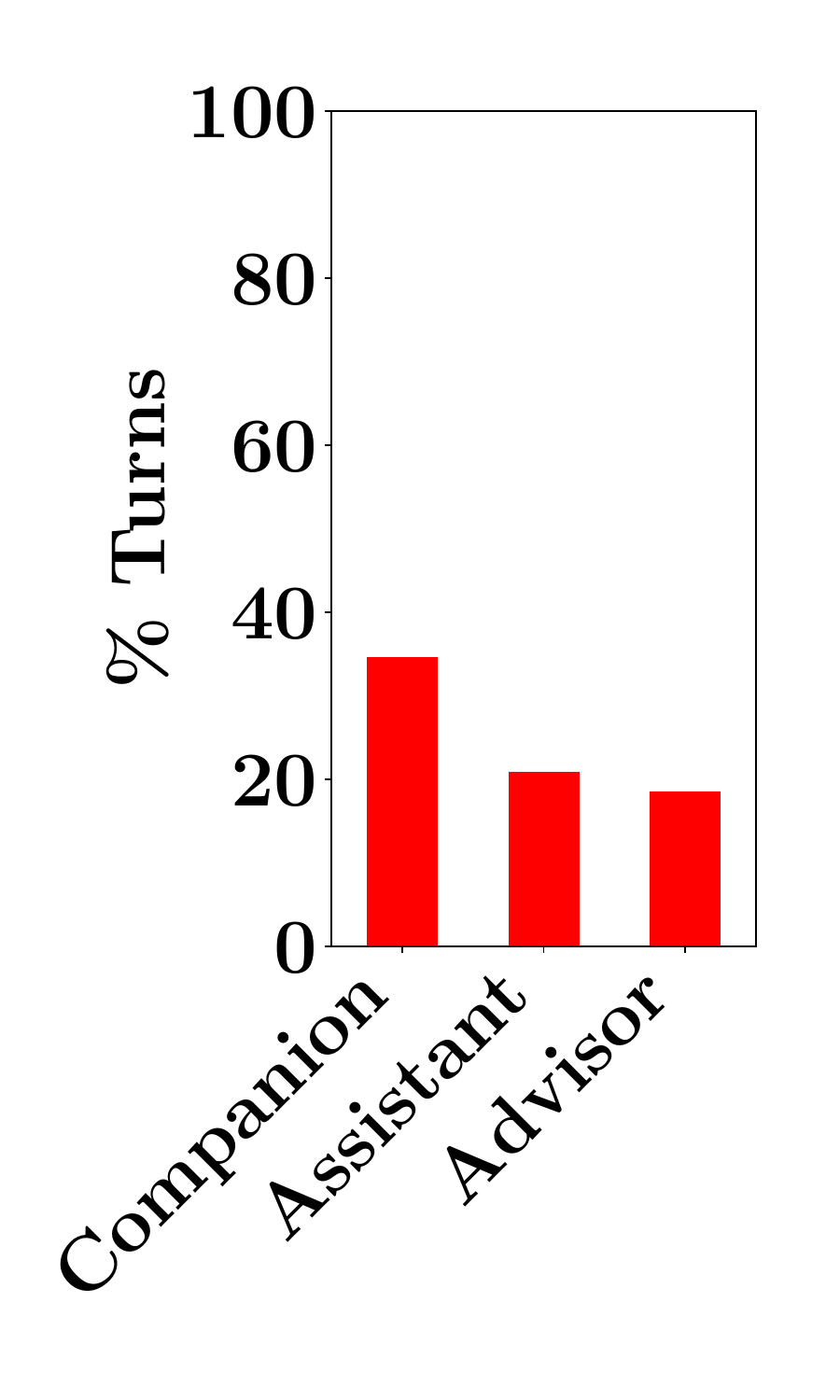}
        \caption{Relationships}
    \end{subfigure}
    \hfill
    \begin{subfigure}{0.69\columnwidth}
        \centering
        \includegraphics[width=\linewidth,height=3.8cm]{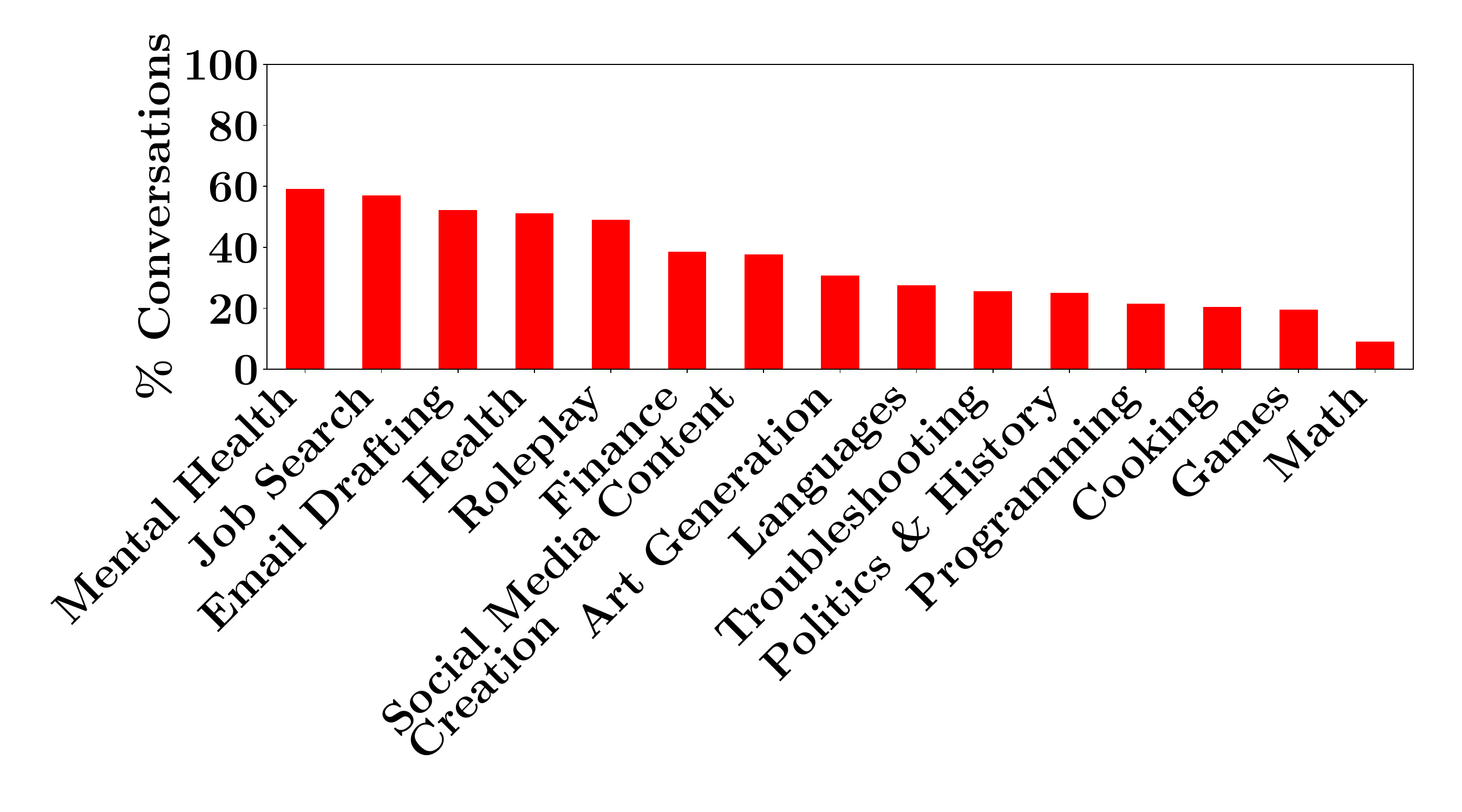}
        \caption{Topics}
    \end{subfigure}
    \caption{\% of conversations/turns that include personal data disclosures across relationships and topics in \gptr{}. Participants divulge more personal information in companion interactions and/or while discussing sensitive topics.
    }
    \label{fig:personal-data-relation-topics}
\end{figure}

Beyond analyzing overall disclosure rates, it is important to understand when and in which context users share personal data. 
To this end, we analyze variation in personal data disclosures across (1)~relationships users attribute to ChatGPT, and (2)~conversation topics. 
We aim to see how the relationship and context shape the likelihood of users divulging personal data.
Figure~\ref{fig:personal-data-relation-topics} reports the percentage of conversation turns that include personal data across relationships and topics. 
%Our results show that the disclosure of personal data is shaped both by the relationships and the topic of discussion. 
Companion interactions stand out as the most prone to disclosure of personal data, with more than \new{35\% of the turns involving participants revealing personal information}, compared to only 18\%-20\% in advisor and assistant.
Personal data disclosures vary by topic: it is especially common in sensitive domains such as mental health, where more than half of conversations involve personal data. 
Similarly, for the health topic, 51\% of the conversations include personal data. Notably, \new{for the topic \textit{Job Search} and \textit{Email Drafting} we observe a significant disclosure of personal data (>50\%) which indicates that people may reveal personal data when asking ChatGPT to undertake some tasks (e.g., disclosing names for email writing or background details for searching jobs).}

% we observe a significant disclosure of personal data (39\%), which indicates that people may reveal personal data when asking ChatGPT to undertake some tasks (e.g., disclosing names for email writing tasks).
%
Our findings highlight that disclosure of personal data is not a uniform behavior; it is conditioned by the relationship and topic, with companion-like roles and sensitive topics creating situations where participants are more prone to revealing personal data. 

% \sk{}
% \am{Anthro vs disclosure?}

\begin{figure}[t!]
    \centering
    \begin{subfigure}[t]{0.45\textwidth}
        \centering
        \includegraphics[width=\linewidth, height = 5cm]{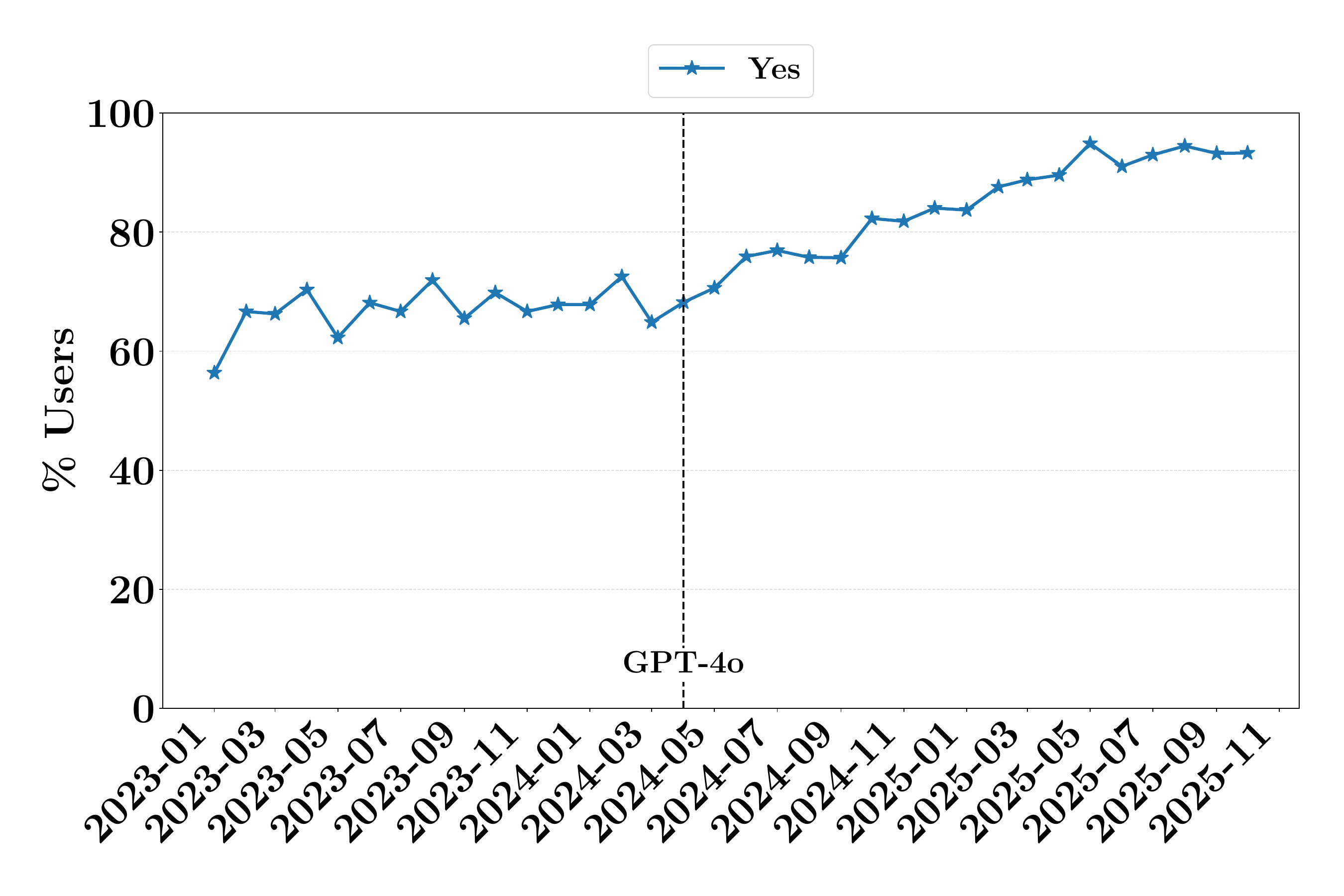}
        \subcaption{\% of participants sharing.}
        \label{fig:percentage-users-personal}
    \end{subfigure}
    \begin{subfigure}[t]{0.45\textwidth}
        \centering
        \includegraphics[width=\linewidth, height = 5cm]{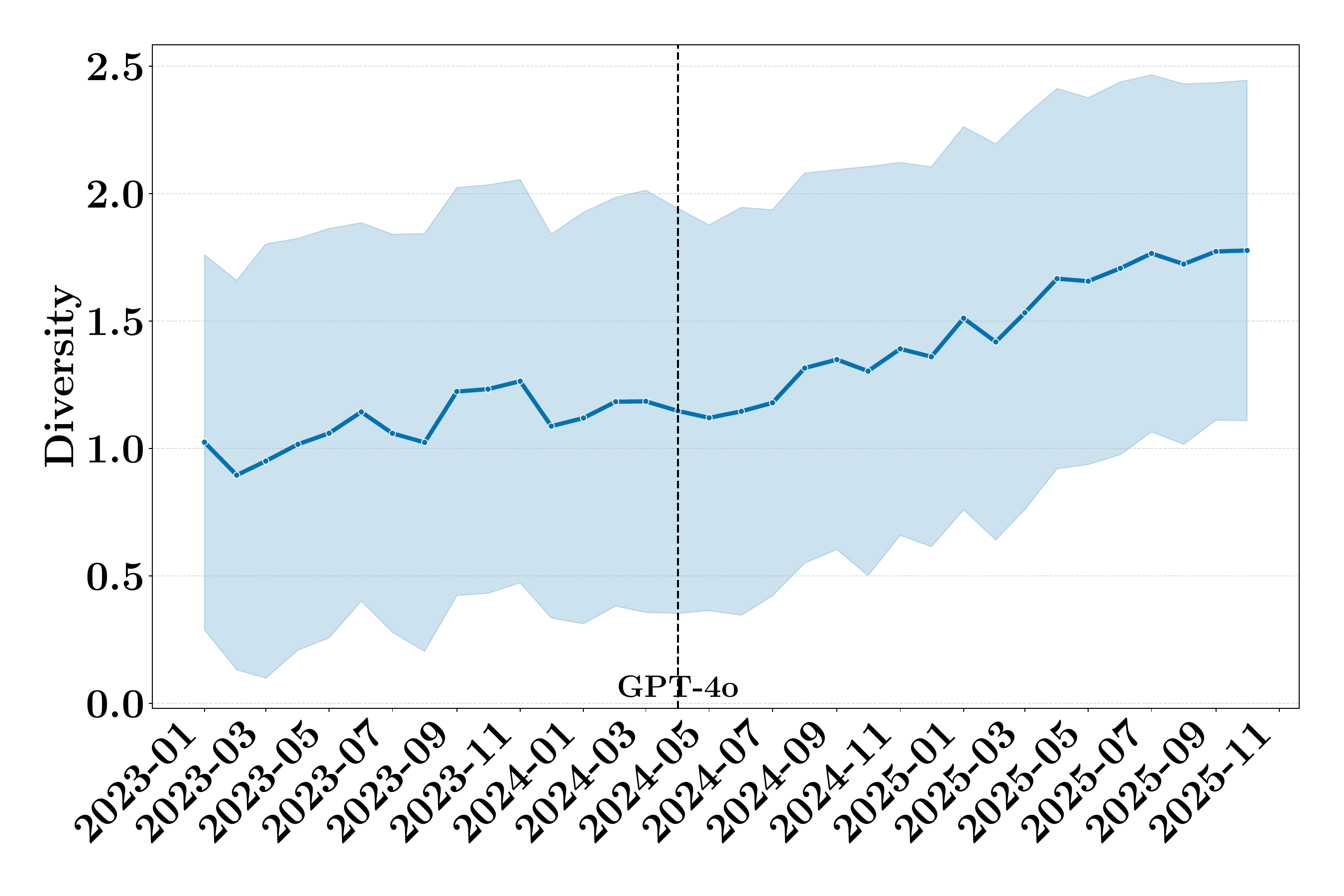}
        \subcaption{Diversity in personal data.}
        \label{fig:diversity-personal-data}
    \end{subfigure}
    \caption{Evolution of personal data sharing in \gptr{}. Increasingly (a)~more and more participants in \gptr{} are sharing (b)~higher diversity of personal data types disclosed with ChatGPT.
    }
    \label{fig:evolution-personal-data}
\end{figure}

\noindent \textbf{Evolution of personal data disclosure.}
While overall disclosure rates reveal where and when personal data is revealed, they do not capture how disclosure behavior evolves as users continue interacting with ChatGPT. 
Here, we aim to study the temporal dimension, which will allow us to observe whether people become more comfortable and disclose more personal data over time.
This perspective is essential for understanding how habits, reliance, and trust in conversational AI systems change over time.
We study this phenomenon by calculating: (1)~the percentage of participants that reveal personal data over time, and (2)~the diversity of different personal data types that are revealed per participant over time (see Figure~\ref{fig:evolution-personal-data}).
%We make the following observations.
We observe that early in the dataset, between 60\% and 70\% of participants disclosed personal information in their conversations. 
After the release of GTP-4o, this percentage steadily increases, reaching over 90\%. 
%This suggests that as the system evolved, participants became more comfortable and likely to reveal personal data. 
Similarly, on the diversity of personal data revealed over time, we observe that in 2023 the diversity is around 1.2 and it has increased to 1.8.

\noindent \textbf{Personal data $\times$ Anthropomorphization}. Users disclosed personal data more frequently when the system exhibits anthropomorphic behavior (26\% of the turns) compared to when it did not (20\% of the turns).

% the number of categories disclosed remained stable, whereas after GPT-4o, users reveal almost twice as many personal data types.
\noindent\textbf{Implications.} Together, these findings suggest that disclosure becomes both more common and diverse over time, pointing to potentially deepening trust in the system, but simultaneously raising concerns about user vulnerability and the adequacy of safeguards for handling sensitive information.

%In order to understand the changes in the types of personal data shared in different topics over time, we look at their distribution in two time ranges - before May 2024 and early 2025 (Figure~\ref{fig:personal-data-vs-topics}). Comparing the two time ranges, we see that in the ``mental health'' topic, people increasingly share information about family, friends, and relationships. Further, the sharing of economic or financial information in the ``finance'' topic nearly doubled over this period. Similarly, the disclosure of mental health related private data has doubled in the ``content generation'' topic.

% \am{By this point readers have forgotten the creamiest part of this section -- anthropomorphism! We need to connect each of the above observations with antropomorphism. In the minimum, how much anthropomorphism in advisor, assistant, companion. How much anthro in different personal data types -- the ones with top anthro?}

\begin{figure}
    \centering
    \begin{subfigure}[t]{0.45\textwidth}
        \centering
        \includegraphics[width=\linewidth, height = 3.5cm]{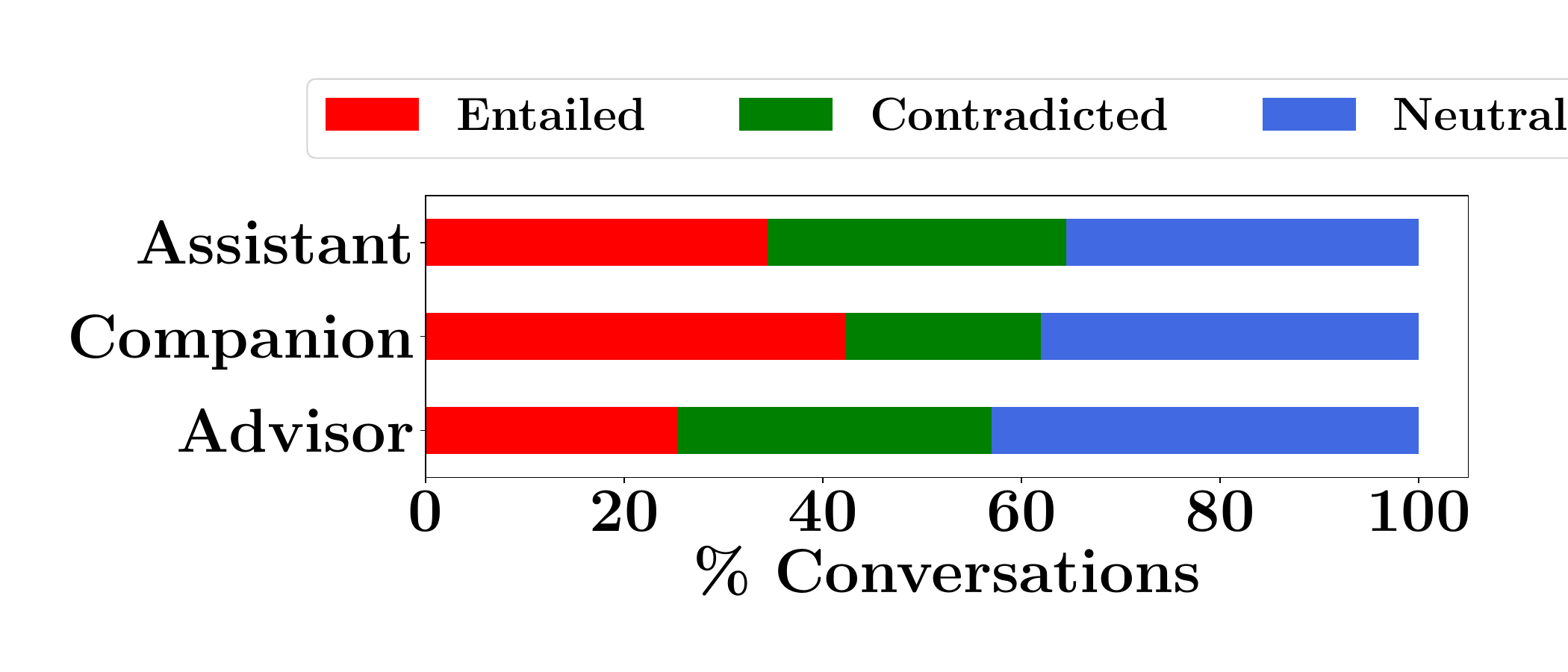}
   % \vspace{-5mm}
    \caption{\label{fig:st-role}Steering across roles. Participants are steered more in conversations where they perceive ChatGPT as a companion.
    }
    \end{subfigure}
    \begin{subfigure}[t]{0.45\textwidth}
        \centering
        \includegraphics[width=\linewidth, height = 3cm]{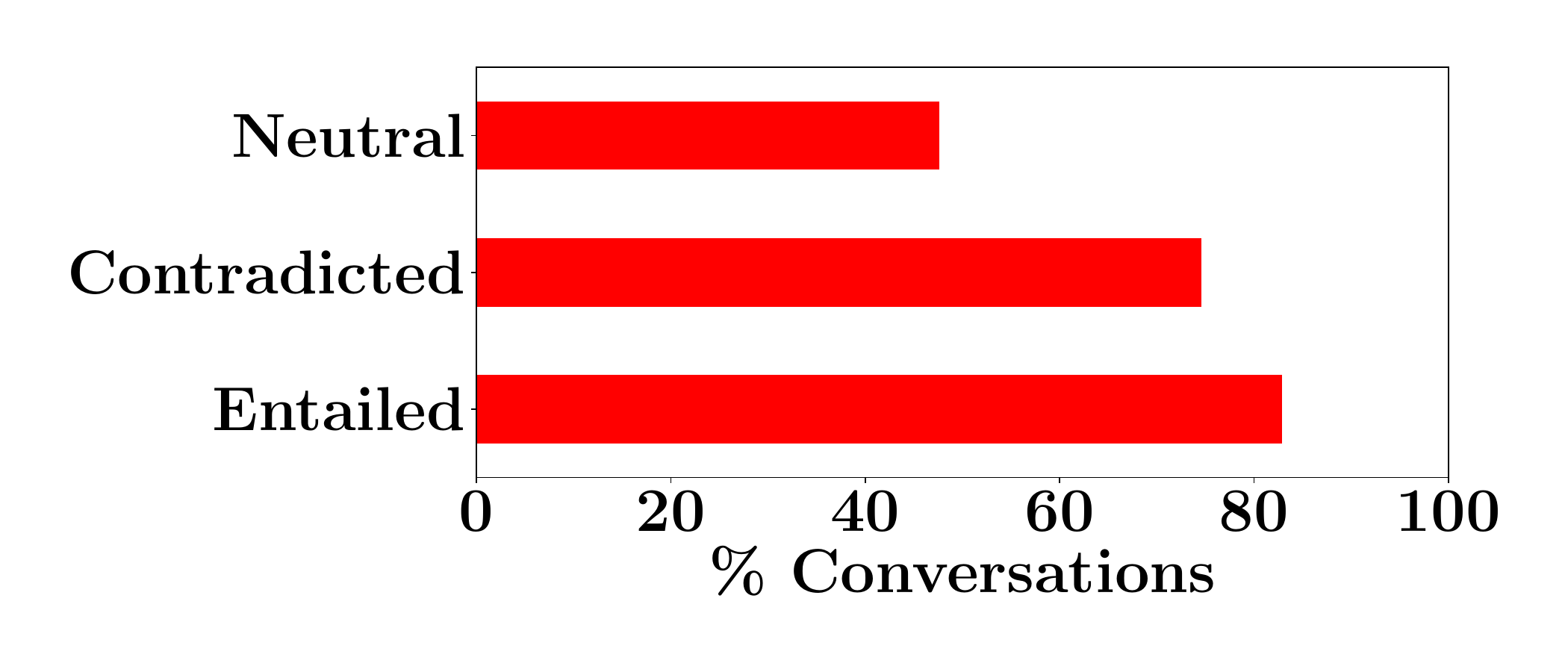}
    %\vspace{-5mm}
    \caption{\label{fig:st-anth}System anthropomorphism in steered conversations. Anthropomorphic behaviour by the system is observed more in entailed conversations.
    }
    \end{subfigure}
    \begin{subfigure}[t]{0.45\textwidth}
        \centering
        \includegraphics[width=\linewidth, height = 3cm]{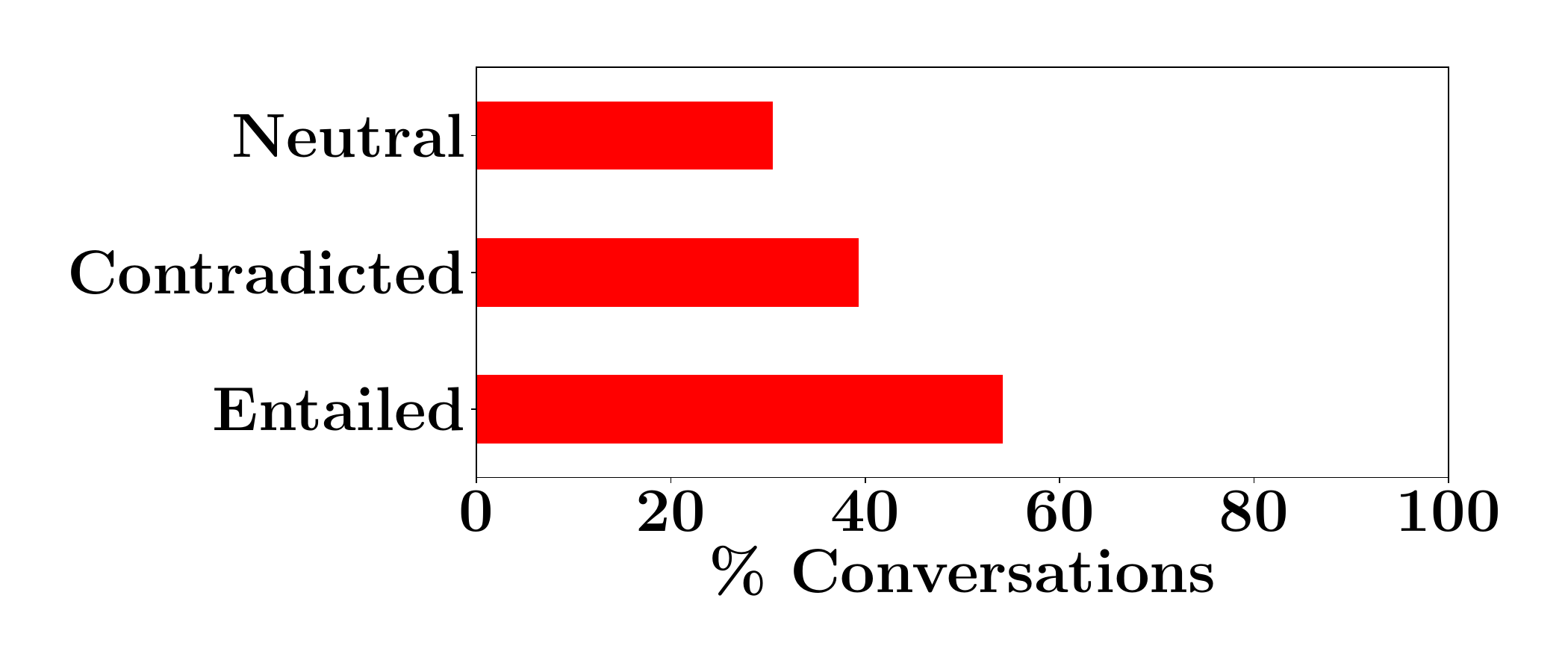}
    %\vspace{-5mm}
    \caption{\label{st-pr}Disclosure of personal data in steered conversations. Participants disclose personal data more frequently in entailed conversations.
    }
    \end{subfigure}
    \caption{Steering versus different attributes.}
    \label{fig:steering-cross-other}
\end{figure}

\begin{figure}[t!]
    \centering

    \begin{subfigure}[t]{0.45\textwidth}
        \centering
        \includegraphics[width=\linewidth, height = 5cm]{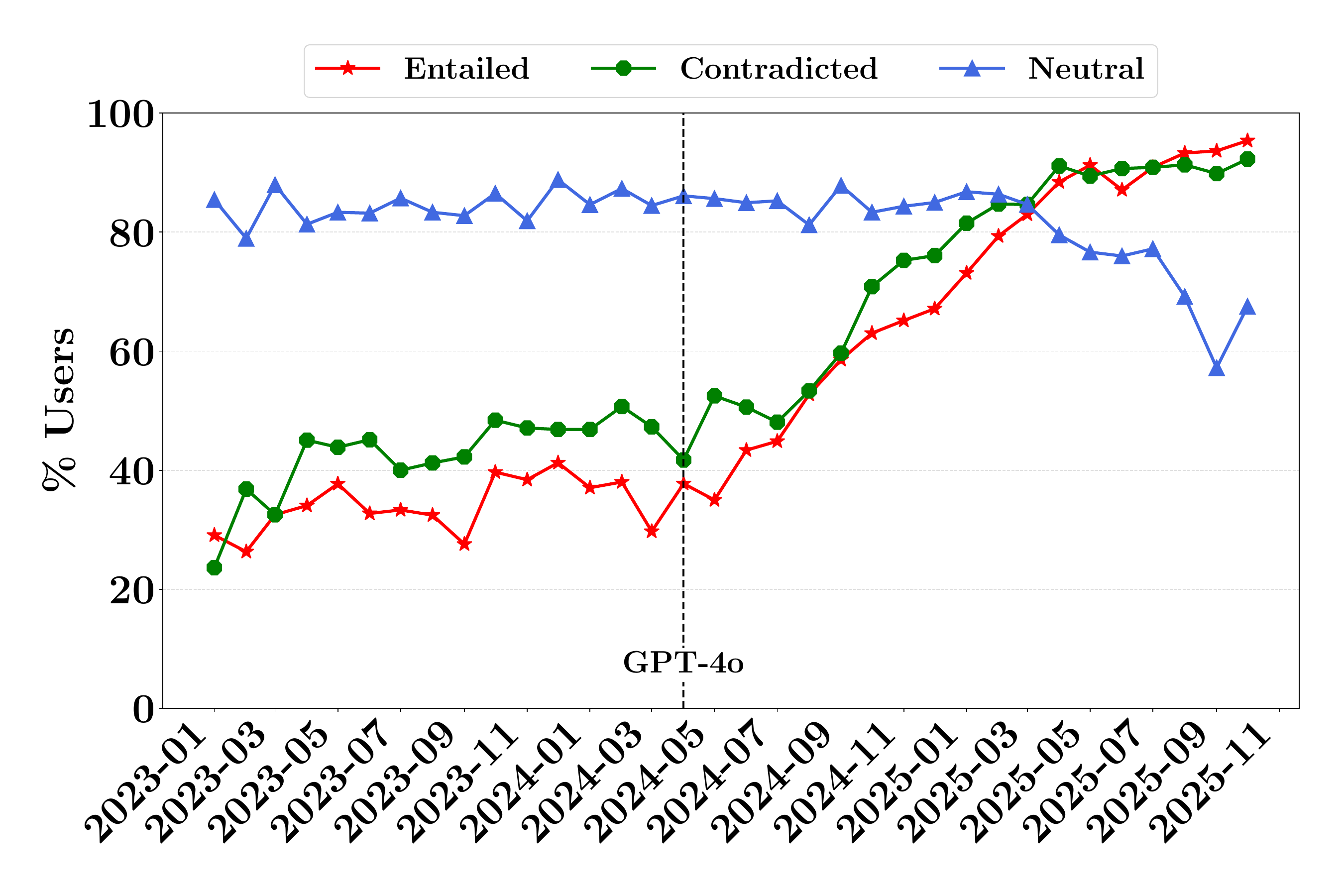}
        \caption{\% of participants with at least one entailed, contradicted, or neutral conversation.}
        \label{fig:evolution-users-steering}
    \end{subfigure}\hfill
    \begin{subfigure}[t]{0.45\textwidth}
        \centering
        \includegraphics[width=\linewidth, height = 5cm]{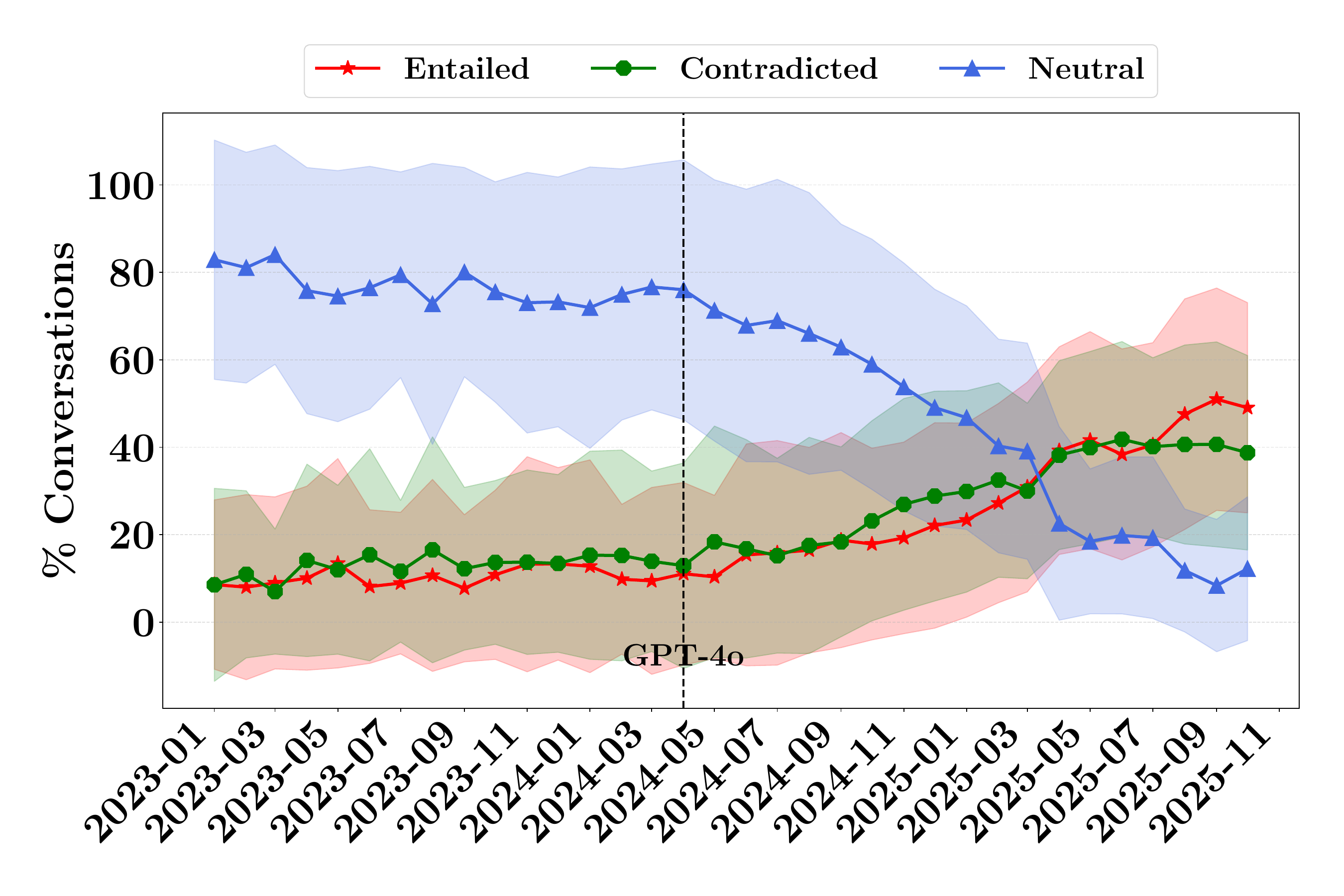}
        \caption{Mean \% of conversation turns per participant.}
        \label{fig:evolution-intensity-users-steering}
    \end{subfigure}\hfill
    \begin{subfigure}[t]{0.45\textwidth}
        \centering
        \includegraphics[width=\linewidth, height = 5cm]{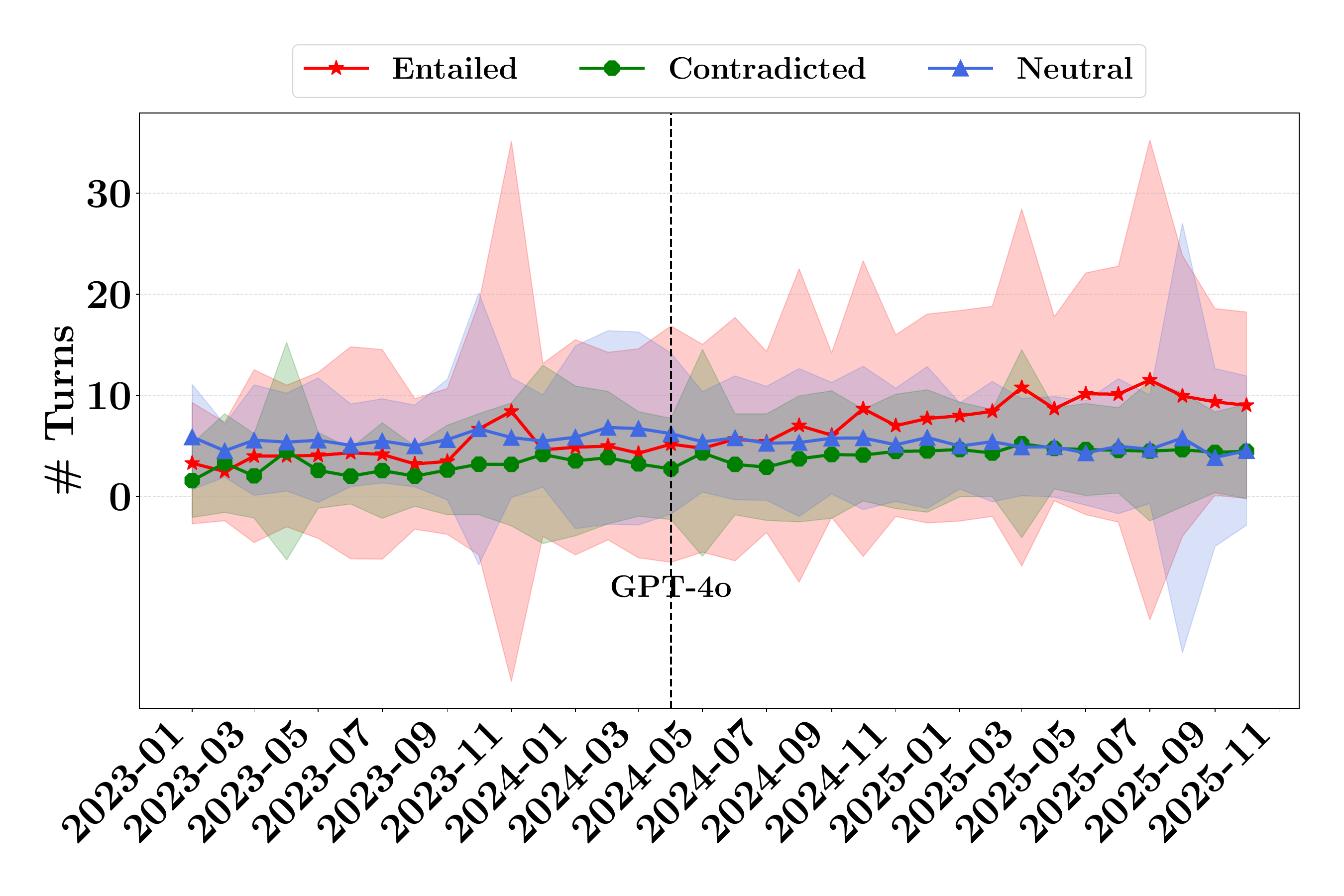}
        \caption{Avg. depth of the conversation per participant.}
        \label{fig:evolution-depth-users-steering}
    \end{subfigure}
    \caption{Temporal evolution of steering in \gptr{}. %After the release of GPT-4o, 
    ChatGPT is increasingly trying to steer conversations with follow-up suggestions and more participants are following model's suggestions (as indicated by growing entailment). 
    }
    \label{fig:evolution-steering}
\end{figure}

%% file: rq3-steering.tex
\section{RQ3 - Conversational steering}
With the growing reliance and evolving relationship between human and AI interactions, next, we analyze conversational steering. We find that \new{18.3\%} of conversational turns are entailed (participants successfully steered by the system), \new{24.6\%} are contradicted (steering attempt failed), and \new{57.1\%} are neutral (no steering attempt). 
At the conversation level, the distribution is \new{30.1\%, 30.2\%, and 39.7\%} for entailed, contradicted, and neutral, respectively.
Figure~\ref{fig:st-role} compares the distribution of steering outcomes across relationships. 
We observe that companion interactions exhibit the highest proportion of entailed conversations (\new{42.3\%}), suggesting that participants in personal or emotionally oriented conversations are more likely to follow model suggestions. 
Assistant conversations show a balanced mix of entailed \new{(34.4\%) and contradicted (30.1\%)} conversations, reflecting a more task-oriented dynamic in which participants selectively accept or reject recommendations. Interestingly, advisor conversations contain the lowest share of entailed conversations \new{(25.3\%)} and the highest rate of contradiction \new{(55.6\%)}, indicating that participants use ChatGPT as an expert for advice. In Figure~\ref{fig:st-anth} and~\ref{st-pr}, we observe that entailed conversations are tied to higher system anthropomorphization ($>80\%$) as well as to higher disclosure of personal data ($\sim 60\%$). Surprisingly, even when contradicting the system's suggestions, participants disclose personal data in approximately $\sim40\%$ of cases.
%Surprisingly, even while contradicting, users tend to reveal personal data in $\sim40\%$ cases.

\noindent \textbf{Temporal evolution.} We study the temporal evolution of conversational steering across three perspectives: (a)~whether users experience steering from the conversational AI system at all, (b)~the prevalence of steering in conversation turns, and (c)~changes in conversation depth over time. 
Figure~\ref{fig:evolution-steering} illustrates our results across these three perspectives.
We observe that a growing share of \new{participants} experience at least one conversation in which the system attempts to steer, and critically, this share increases after the release of GPT-4o (Figure~\ref{fig:evolution-users-steering}). 
In particular, after GPT-4o, the proportion of \new{participants} with at least one entailed conversation steadily increases, converging with the proportion of \new{participants} with contradicted conversations by early 2025. 
This result indicates both the rise in conversational steering and that more \new{participants} are following the model's suggestions.

These results are reinforced when looking at the prevalence of entailed, contradicted, and neutral conversation turns (Figure~\ref{fig:evolution-intensity-users-steering}) over time. 
The mean percentage of turns that are entailed exhibits a shift following GPT-4o. 
Before the release, only \new{11\%} of the conversations contained a successfully steered turn. 
After, GPT-4o this proportion increases to almost \new{50\%}, suggesting that the updated model initiates follow-up suggestions more frequently and \new{participants} are more likely to respond in-line with the model's suggestions. 
Neutral conversations remain the majority throughout the period, but their dominance declines post-GPT-4o release. % as both entailed and contradicted turns grow. 
This pattern reflects a transition from a passive conversational model toward one that is more active, as it introduces continuations and follow-ups.
Finally, we sought to determine whether steering might enhance user engagement by increasing the depth of conversations. 
We find that indeed this is the case (see Figure~\ref{fig:evolution-depth-users-steering}): for conversations that do involve steering, depth increases especially after GPT-4o. 
This correlation suggests that successful steering fosters deeper and more extended interactions.
%This suggests that successful steering is associated with deeper and more extended interactions. %, whereas failed or absent steering does not correlate with increased conversation depth.

\noindent
\textbf{Implications. } Our findings suggest that newer models are 
% not only improving performance but also 
reshaping the conversational dynamic of human-AI interactions, as systems shift from reactive responders toward more proactive agents. As steering becomes more frequent and more accepted by users, we must consider how model-initiated suggestions may influence user autonomy, especially when such behavioral shifts arise from backend updates that users may not notice. 
Altogether, there is a need for more transparency and guardrails to ensure that proactive models support, rather than inadvertently steer, users in possibly unintended directions.

%\sz{i think it will be interesting to check for the entailed or contradicted cases, whether we find instances where the model tries to steer the user to a direction that is quite unrelated to the topic of the conversation. probably an idea for a different paper to study more indepth these "nudges" and user responses}

%\sz{Flow in paper:
%Analyze who is steering the conversation in general and over time
%Analyze differences across topics and relationships
%Qualitative insights into how this steering happens in practice
%}

% \begin{itemize}
%     \item Distribution of conversation turns - 9.6\% Entailment, 22.8\% Contradiction and 67.6\% Neutral.
%     \item Distribution of conversations - 25.1\% Entailment, 27.4\% Contradiction and 47.5\% Neutral.
%     \item no variation is observed across topics, disclosure of personal data.
% \end{itemize}

%% file: Limitation.tex
\section{Limitations}\label{appendix:ethics_limitations}
% \noindent \textbf{Ethical considerations:} 
% \new{This study is conducted with careful attention to ethical considerations and is approved by the Ethical Review Board (ERB) of our university. %Saarland University. % to conduct this research.
% Participants were first informed about the risks associated with sharing their ChatGPT data, and secondary usage of NLP tools to analyze the conversations. 
% Thereafter, with explicit and informed consent they donated us their ChatGPT data exports. 
% The donated datasets are stored on secure servers with strict access control. 
% The data can neither be shared with any third party nor will be released publicly due to their inherent sensitive nature.
% As per the ERB guidelines we will delete the data within 3 years of completion of this study. 
% For analyses conducted in the paper, we rely on GPT-4o as a judge to scale up annotations. 
% The choice is motivated by two important considerations: (a)~GPT-4o is a model from OpenAI and is the model with whom participants had originally had the conversations, (b)~we leverage the EDU workspace of OpenAI which provides strict data protection and restrictions for usage of this data for training OpenAI models.} 
% % All the used open-source language models are deployed on the institute's secured servers that abide by rigorous data protection and access control principles.

Our study has two main limitations. First, while our dataset is quite rich, it includes a relatively small number of users (300). 
However, the goal of this work was not to produce exact population-level estimates but to uncover trends and highlight emerging risks in the evolution of human–AI interactions. The insights we provide should therefore be read as indicative of broader dynamics rather than as precise measurements. 
Second, our annotations relied on GPT-4o to classify topics, anthropomorphization, relationships, personal data and steering. Although this approach enabled scalable and consistent analysis, the model is not perfect and may mislabel some cases. We mitigated this limitation by validating subsets of the annotations, but future work should strengthen this process through complementary annotation strategies.

%% file: Conclusion.tex
\section{Concluding discussion}~\label{Sec : Dscussion}
%\todo{We need to cite the memories paper highlighting other utilities of this dataset \cite{dash2026memories}}
In this work, we studied how human–AI interactions evolved over time by analyzing
the \gptr{} dataset with ChatGPT traces donated by 300 users. 
%ChatGPT data donated by \new{300} users.
Our longitudinal analysis reveals clear shifts in \emph{purpose}, \emph{framing}, and \emph{steering}. 
\new{Participants} expand the purposes for which they turn to ChatGPT, increasingly consulting it about a broader and more sensitive range of topics such as health and \new{mental health}.
At the same time, the framing of interactions changes: while ChatGPT continues to serve as an assistant or advisor, its role as a companion grows sharply, spreading across more \new{participants}, occurring more frequently, and developing into deeper conversations. 
Mirroring this, personal data disclosures become widespread and more diverse, particularly in the companion mode.  
Finally, we observe a rise in conversational steering after the release of GPT-4o, with the system increasingly proposing directions that the \new{participants} choose to follow.
Our findings have important implications for multiple stakeholders. For \emph{users}, our results highlight both opportunities and risks of treating ChatGPT as more than a functional tool, underscoring the need to remain mindful of when and how sensitive information should be disclosed.
For \emph{AI designers}, the growing prevalence of companion-like conversations and model-initiated steering highlights the need for safeguards that support user autonomy, such as mechanisms for setting conversational boundaries. % and reminders for system limitations. 
For \emph{policymakers}, our work highlights the need for policies that address the relational and evolving nature of AI use to ensure user autonomy and privacy are protected as these systems become more embedded in everyday life. %Measures could include clear disclosure cues, stronger safeguards for sensitive topics, and oversight mechanisms to protect user autonomy and privacy as these systems become increasingly embedded in everyday life.
%For instance, requiring clear disclosure cues, stronger safeguards for sensitive topics, and oversight mechanisms that ensure user autonomy and privacy are protected as these systems become more embedded in everyday life.

% \begin{enumerate}
%     \item Summary of observations
%     \item Limitations
%     \item Takeaways for stakeholders
%         \begin{enumerate}
%             \item Users
%             \item AI designers
%             \item Policymakers
%         \end{enumerate}
% \end{enumerate}